\documentclass[a4paper,11pt]{article}
\usepackage{graphicx}
\usepackage{caption}
\usepackage{subfig}
\usepackage{bm}
\usepackage{verbatim}
\usepackage{microtype}
\usepackage{amsmath}
\usepackage{siunitx}
\usepackage{physics}
\usepackage{amssymb}
\usepackage{slashed}
\usepackage{pos}
\usepackage{hyperref}
\newcommand{\bea}{\begin{eqnarray}}
\newcommand{\eea}{\end{eqnarray}}
\newcommand{\bs}{\boldsymbol}

\title{Virtual Photon Emission in Leptonic Decays of Pseudoscalar Mesons}
\author[a]{R. Frezzotti}
\author[b]{G. Gagliardi}
\author[b]{V. Lubicz}
\author[c]{G. Martinelli}
\author*[b]{F. Mazzetti}
\author[d]{C. Sachrajda}
\author[e]{F. Sanfilippo}
\author[e]{S. Simula}
\author[a]{N. Tantalo}

\affiliation[a]{Dipartimento di Fisica and INFN, Università di Roma Tor Vergata,
Via della Ricerca Scientifica 1, Rome, Italy}

\affiliation[b]{Dipartimento di Matematica e Fisica and INFN, Università di Roma Tre, Via della Vasca Navale 84, Rome, Italy}

\affiliation[c]{Dipartimento di Fisica and INFN, Università di Roma La Sapienza, Piazzale Aldo Moro 5, Rome, Italy}

\affiliation[d]{Department of Physics and Astronomy, University of Southampton, SO17 1BJ, UK}

\affiliation[e]{Istituto Nazionale di Fisica Nucleare, Sezione di Roma Tre, Via della Vasca Navale 84, Rome, Italy}

\emailAdd{filippo.mazzetti@uniroma3.it}

\abstract{We present a preliminary non-perturbative lattice calculation of the form factors entering the processes $K\to \ell\,\nu_\ell\,\ell'^+\,\ell'^-$ and of the corresponding branching ratios. These form factors describe the interaction between the mediating virtual photon and the internal hadronic structure of the meson. By separating them from the point-like contribution to the matrix element we are able to isolate and reconstruct the structure-dependent contribution to the decay width. Our numerical analysis employs only one gauge ensemble and so it is affected by systematic uncertainties due to the missing continuum and physical point extrapolation. Despite this, we already find a reasonable agreement with the experimental data and with the next to leading order Chiral Perturbation Theory predictions. 
The method is general and can be applied to any pseudoscalar meson, though for heavier mesons the possibility of internal lighter states becomes problematic and still needs a proper study.
A non-perturbative, model-independent lattice evaluation of these processes would allow further progress in the theoretical predictions of SM hadronic quantities and in the search of New Physics. }

\FullConference{%
 The 38th International Symposium on Lattice Field Theory, LATTICE2021
  26th-30th July, 2021
  Zoom/Gather@Massachusetts Institute of Technology
}

\begin{document}
\maketitle

\section{Introduction}
$P\to \ell\,\nu_\ell\,\ell'^+\,\ell'^-$ decays, where $P$ is a pseudoscalar meson, are suppressed processes, the width of which starts at the second order in $\alpha_{em}$. For this reason, a precise Standard Model (SM) theoretical prediction is a crucial step, together with precise experimental measurements, to identify small hypothetical New Physics (NP) effects. 
From the theoretical point of view, the computation of the decay rate requires the knowledge of four hadronic form factors, that depend on the invariant masses of the two leptonic pairs $\ell\,\nu_\ell$ and $\ell'^+\,\ell'^-$. They describe the interaction between the virtual photon and the internal hadronic structure of the decaying meson. These structure dependent (SD) form factors, and their contribution to the decay rate, are the only actually relevant quantities that need to be evaluated and compared to the experiments. Indeed, the contributions coming from the lepton Bremsstrahlung and from the meson point-like approximation are already well known and the meson decay constant is the only hadronic parameter needed to evaluate them. 
At present, for kaon decays the theoretical knowledge of the SD form factors comes from Chiral Perturbation Theory (ChPT), which has been used at next to leading order (NLO) to estimate their value and their contribution to the branching ratios in \cite{Bijnens:1994me}. It is worth noticing that at NLO order in ChPT form factors are constants. For heavy mesons ChPT does not apply, and the one theoretical prediction is presented in \cite{Danilina_2020} for $B$ decays, where a Vector Meson Dominance model has been used. The prediction for the $B^+\to\mu^+\,\nu_\mu\,\mu^+\,\mu^-$ branching ratio of \cite{Danilina_2020}, however, is almost four times larger than the experimental upper limit obtained in \cite{Aaij_2019}.
In this context, it is clear that a non-perturbative, model independent lattice evaluation of the SD form factors is essential. For kaon decays it would allow to test the ChPT predictions, to study the momentum dependence of the form factors and to increase the precision with which they are known now. For heavy mesons a lattice computation is even more necessary, since we can not rely on ChPT. 
At the present, the only lattice study of these processes has been recently presented in \cite{xu}, where a method to compute $K\to \ell\,\nu_\ell\,\ell'^+\,\ell'^-$ has been applied to derive the whole branching ratio for different final leptons. Even if the analysis in \cite{xu} is based only on one ensemble, with unphysical pion mass, their result are already close to the experimental data and ChPT predictions. However the method implemented in \cite{xu} does not provide values for the SD form factors and also the SD contribution to the branching ratios is not separated from the point-like contribution. 
We propose a new strategy to study on the lattice $P\to \ell\,\nu_\ell\,\ell'^+\,\ell'^-$ decays, and we apply it to perform an exploratory study of kaon decays by using one of the $N_f=2+1+1$ gauge
ensembles generated by the European Twisted Mass Collaboration. Our method allows to estimate individually each of the SD form factors entering the amplitude, and to study their dependence on the kinematic variables. With them one can reconstruct separately all the contributions to the branching ratios, namely the purely SD one, the point-like one, and the one coming from the interference between the two. Despite the method being completely general and in principle applicable to any pseudoscalar meson, in practice for heavy mesons some complications arise when the virtuality of the off-shell photon becomes larger then twice the pion mass, namely in the region $k^2>4m_\pi^2$, where $k$ is the virtual photon four-momentum. For the present case the unphysical pion mass, such that $m_K<2 m_\pi$, prevents us from encountering such complications (this is also the case in the analysis presented in \cite{xu}). This issue, together with a complete study of all the systematic effects (due to discretization, finite volume and unphysical quark masses) will be object of future works.

\section{The Hadronic Matrix Element From Lattice Correlator}
\begin{figure}
	\begin{center}
	\subfloat{%
		\includegraphics[scale=0.15]{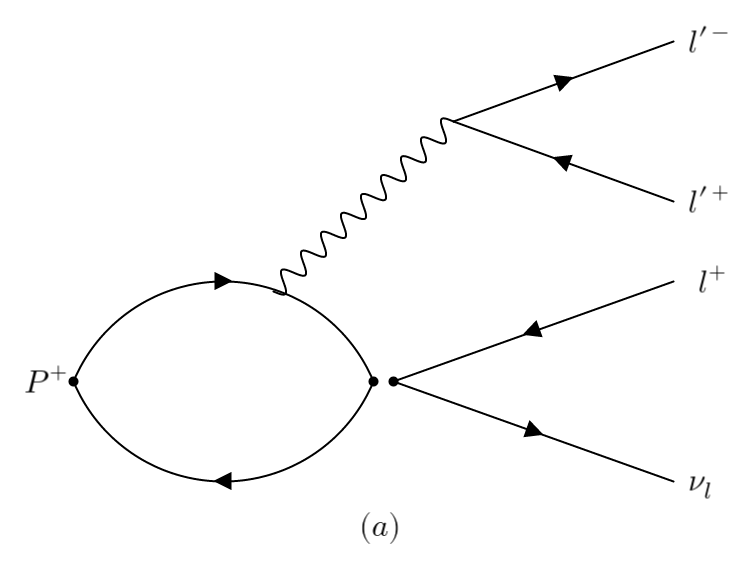}}
	\subfloat{%
		\includegraphics[scale=0.15]{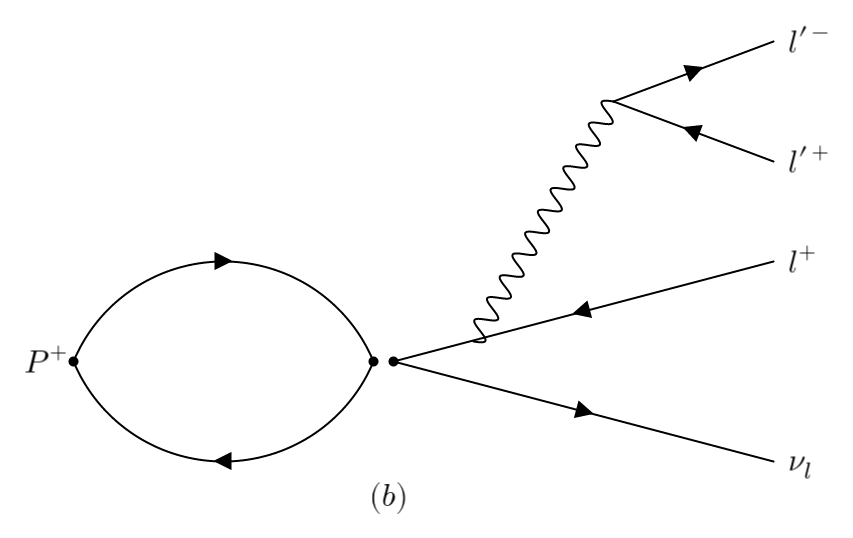}} 
	\caption{\it Diagrams contributing to the process $P^+\to l^+\,\nu_l\,l'^+\,l'^-$.}
	\label{feynman}
	\end{center}
\end{figure}

At lowest order in the electroweak interaction, $P^+\to \ell^+\,\nu_\ell\,\ell'^+\,\ell'^-$ decays are obtained from the diagrams depicted in Fig. \ref{feynman}. If $\ell=\ell'$, we also need to consider the diagrams obtained by interchanging the two identical charged leptons.
The diagram \ref{feynman}(b) can readily be computed in perturbation theory, with the meson decay constant as the only required non-perturbative input. In diagram \ref{feynman}(a) the non-perturbative hadronic contribution to the matrix element factorizes, and is encoded in the following tensor:
\bea\label{tensor}
H_W^{\mu\nu}(k,p)=\int d^4x e^{ik\cdot x}\mel{0}{T[J_{\mathrm{em}}^\mu(x) J_W^\nu(0)]}{P(p)}\label{hadronic}\,,
\eea
where $J_{\mathrm{em}}^\mu(x)$ is the electromagnetic current, $J_W^\nu$ is the weak current, $k=(E_\gamma, \boldsymbol{k})$ is the four-momentum of the virtual photon and $p=(E, \boldsymbol{p})$ is that of the incoming pseudoscalar meson $P$. The meson and photon energies satisfy $E=\sqrt{m_P^2+\boldsymbol{p}^2}$ and $E_\gamma=\sqrt{k^2+\boldsymbol{k}^2}$.
The hadronic tensor can be decomposed into form factors which are scalar functions encoding the non-perturbative strong dynamics. Following Ref.\,\cite{Carrasco_2015}, we write:
\bea
H_W^{\mu\nu}&=&H^{\mu\nu}_{\mathrm{pt}}+H^{\mu\nu}_{\mathrm{SD}}\,,\quad\quad\quad
H^{\mu\nu}_{\mathrm{pt}}=f_P\left[g^{\mu\nu}-\frac{(2p-k)^\mu(p-k)^\nu}{(p-k)^2-m_P^2}\right]\,,\nonumber\\
H^{\mu\nu}_{\mathrm{SD}}&=&\frac{H_1}{m_P}\left(k^2g^{\mu\nu}-k^\mu k^\nu\right)+\frac{H_2}{m_P}\frac{\left[(k\cdot p-k^2)k^\mu-k^2\left(p-k\right)^\mu\right]}{(p-k)^2-m_P^2}\left(p-k\right)^\nu\nonumber\\& &+\frac{F_A}{m_P}\left[(k\cdot p-k^2)g^{\mu\nu}-(p-k)^\mu k^\nu\right]
-i\frac{F_V}{m_P}\epsilon^{\mu\nu\alpha\beta}k_\alpha p_\beta\,.\label{Hmunu_sd}
\eea
With this decomposition we separate the point-like contribution to the hadronic tensor from the structure-dependent one. The former depends only on the meson decay constant and is obtained by assuming a point-like meson. Instead, the SD contribution describes the interaction between the virtual photon and the hadronic structure of the pseudoscalar meson. The SD form factors, $H_1$, $H_2$, $F_A$ and $F_V$, are infrared finite and scalar functions of $k^2$ and $(p-k)^2$.
The main goal of this lattice study is to compute the SD form factors in order to reconstruct the full matrix element and subsequently the branching ratio for the decay. We do this in a way which separates the point-like contribution from that which depends on the hadronic structure. 

\begin{figure}[!t]
	\begin{center}
		\includegraphics[width=0.425\textwidth]{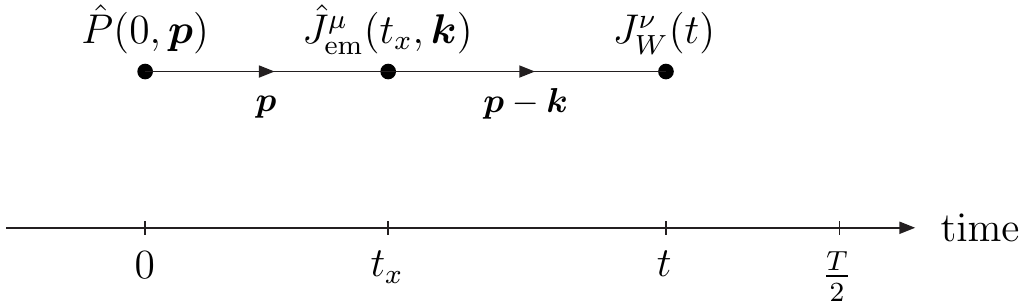}\qquad\qquad
		\includegraphics[width=0.425\textwidth]{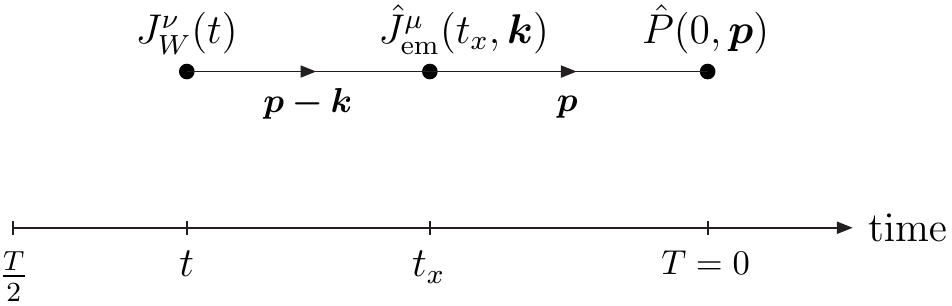}
	\caption{\it{Schematic diagrams representing the correlation function $C_W^{\mu\nu}(t,  E_\gamma,\boldsymbol k,\boldsymbol p)$ used to extract the form factors, see Eqs.\,(\ref{eq:MW}) and (\ref{eq:Cmunudef}).  The interpolating operator for the meson $\hat{P}$ and the weak current $J_W$  are placed at fixed times $0$ and $t$ and the electromagnetic current $\hat{J}_{\mathrm{em}}$ is inserted at $t_x$ which is integrated over $0 \le t_x\le T$, where $T$ is the temporal extent of the lattice. The left and right panels correspond to the leading contributions to the correlation function for $t_x<T/2$ and $t_x>T/2$ respectively, with mesons propagating with momenta $\bs{p}$ or $\bs{p}-\bs{k}$. } \hspace*{\fill}
	\label{fig:correlator}}
	\end{center}
\end{figure}

On the lattice, correlation functions can only be computed in Euclidean space-time, thus we need to translate the Minkowski Green function to the corresponding Euclidean one by employing the analytical continuation to Euclidean time (also called Wick rotation). This operation can be done without difficulties depending on the analytic structure of the $T$-product in Eq.\,(\ref{tensor}):
the presence of singularities (poles or cuts) in Minkowski space can prevent the possibility of making a naive Wick rotation. The presence of such singularities implies the existence of intermediate states with energies which are smaller than the external ones, resulting in integrals over the Euclidean time which grow exponentially with the upper limit of integration (the lattice time extension)\cite{maiani}. By performing a spectral analysis of the matrix element in Eq.\,(\ref{tensor}) we found that internal lighter states do occur when $k^2>4m_\pi^2$, while for $k^2<4m_\pi^2$ the Wick rotation can be performed without any obstacle. For now we restrict our numerical analysis to kaon decays with an unphysical pion mass such that $m_K<2m_\pi$, so that the condition $k^2<4m_\pi^2$ is always satisfied. In the future we will address the problem in more detail, so to extend the method also to heavier mesons. 
\\
On the lattice, with finite space-time volume $V=L^3\times T$, we compute the following three-point correlation function (see also Fig. \ref{fig:correlator}):
\bea\label{eq:correlator22}
& &C_W^{\mu\nu}(t,E_\gamma, \boldsymbol{k}, \boldsymbol{p})=-i\theta\left(T/2-t\right)\sum^T_{t_x=0}\left(\theta\left(T/2-t_x\right)e^{E_\gamma\,t_x}+\theta\left(t_x-T/2\right)e^{-E_\gamma(T-t_x)} \right)M_W^{\mu\nu}(t_x,t;\boldsymbol{k},\boldsymbol{p})\nonumber\\
& &-i\theta\left(t-T/2\right)\sum^T_{t_x=0}\left(\theta\left(T/2-t_x\right)e^{-E_\gamma\,t_x}+\theta\left(t_x-T/2\right)e^{-E_\gamma(t_x-T)} \right)M_W^{\mu\nu}(t_x,t;\boldsymbol{k},\boldsymbol{p})\,.\label{eq:Cmunudef}
\eea
where
\bea\label{eq:MW}
M_W^{\mu\nu}(t_x,t;\boldsymbol{k},\boldsymbol{p})=T\langle J^\nu_W(t)\hat{J}^\mu_{\mathrm{em}}(t_x,\boldsymbol{k})\hat{P}(0,\boldsymbol{p})\rangle_{LT}\,,
\eea
and $\langle ... \rangle_{LT}$ denotes the average over the gauge field configurations at finite L and T.
In the previous formula $\hat{P}(0,\boldsymbol{p})$ is the spatial Fourier transform of the interpolating operator for the decaying pseudoscalar meson at time $t=0$, while the renormalised  hadronic weak current, $J_W^\nu(t)= J_V^\nu(t)-J_A^\nu(t)$ is placed at a generic time $t$ and at the origin in space. Finally we have the electromagnetic current $\hat{J}^\mu_{\mathrm{em}}(t_x, \boldsymbol k)$ which carries spatial momentum $\boldsymbol k$. All details about the lattice discretization of the operators that we employed in our numerical analysis can be found in \cite{Desiderio2020} where the case of real photon emission has been addressed.
Finally we mention that in our analysis we use the so-called electroquenched approximation in which the sea-quarks are electrically neutral.

In order to show that it is possible to extract the hadronic matrix element in Eq.\,(\ref{tensor}) from the function in Eq.\,(\ref{eq:correlator22}) we perform a spectral decomposition of $C_W^{\mu\nu}(t)$.
On the assumption that $k^2<4m_\pi^2$, we derive the relation
	\bea\label{C-H}
	C_W^{\mu\nu}(t,E_\gamma,\boldsymbol{k},\boldsymbol{p})&=& \Theta(T/2-t)\,\frac{e^{-t(E-E_\gamma)}\, \langle P\vert P\vert  0\rangle}{2E}\,H_W^{\mu\nu}(k,\boldsymbol p)\nonumber\\ & &+\ \Theta(t-T/2)\,  \frac{e^{-(T-t)\, (E-E_\gamma)}\langle 0\vert  P\vert P\rangle }{2E}\,\left[H_W^{\mu\nu}(k,\boldsymbol p)\right]^\dagger+\ \dots
	\eea
where the dots represent the sub leading exponentials, suppressed as $e^{-\Delta E t}$ or $e^{-\Delta E(T-t)}$, where $\Delta E$ is the positive energy gap between the internal states contributing to the correlator and the external ones.
We also see that when $t>T/2$ the correlator represents the time-reversal of the original process.
It is useful to note that, in order to separate the axial and vector form factors, it is enough to compute separately the correlation functions corresponding to the vector, $C^{\mu\nu}_V(t, E_\gamma,\boldsymbol k,\boldsymbol p)$, and the axial, $C^{\mu\nu}_A(t, E_\gamma,\boldsymbol k,\boldsymbol p)$, components of the weak current. Moreover, from the properties
\bea
\left[H^{\mu\nu}_A(k,p)\right]^\dagger=H_A^{\mu\nu}(k,p),\quad \left[H_V^{\mu\nu}(k,p)\right]^\dagger=-H_V^{\mu\nu}(k,p)
\eea
we deduce the following properties of the corresponding correlation functions under time reversal:
\bea
C^{\mu\nu}_{A}\left(T-t, T/2, E_\gamma, \boldsymbol{k}, \boldsymbol{p}\right)&=&C^{\mu\nu}_{A}\left(t, T/2, E_\gamma, \boldsymbol{k}, \boldsymbol{p}\right),\nonumber\\ C^{\mu\nu}_{V}\left(T-t, T/2, E_\gamma, \boldsymbol{k}, \boldsymbol{p}\right)&=&-C^{\mu\nu}_{V}\left(t, T/2, E_\gamma, \boldsymbol{k}, \boldsymbol{p}\right)\,.
\eea 
We use these time reversal properties of the lattice correlators, to either symmetrize or anti-symmetrize the correlators between the two halves $[0, T/2]$ and $[T/2,T]$ of the lattice and then we will work just within the first half of the lattice time-extent, where we define
\bea\label{eq:lat_hadr_tens}
H^{\mu\nu}_L(t,k,\boldsymbol p)=\frac{2E}{e^{-t(E-E_\gamma)}\mel{P}{P}{0}} C_W^{\mu\nu}(t,E_\gamma,\boldsymbol{k},\boldsymbol{p})= H_W^{\mu\nu}(k,\boldsymbol p)+...
\eea
where the subscript $L$ stands for "lattice" and the ellipsis represents the sub-leading exponentials.

In this section we have shown how to obtain the hadronic tensor from the lattice correlation function and we now proceed to discuss the extraction of all the SD hadronic form factors and the numerical results we obtained.

\section{Numerical Results for the SD Form Factors}\label{sdff}
\begin{figure}
	\begin{center}
	\subfloat{%
		\includegraphics[scale=0.30]{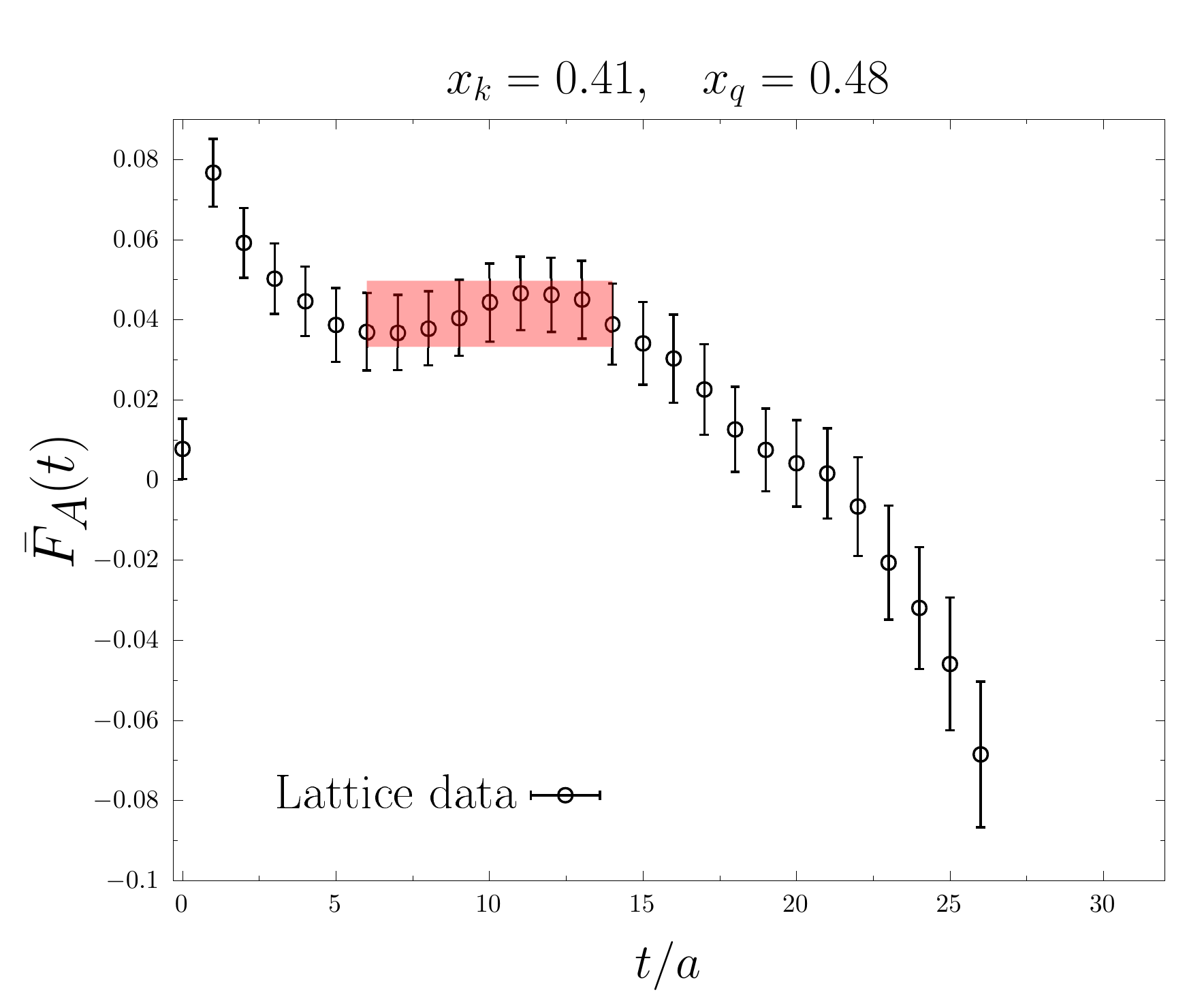}}
	\subfloat{%
		\includegraphics[scale=0.30]{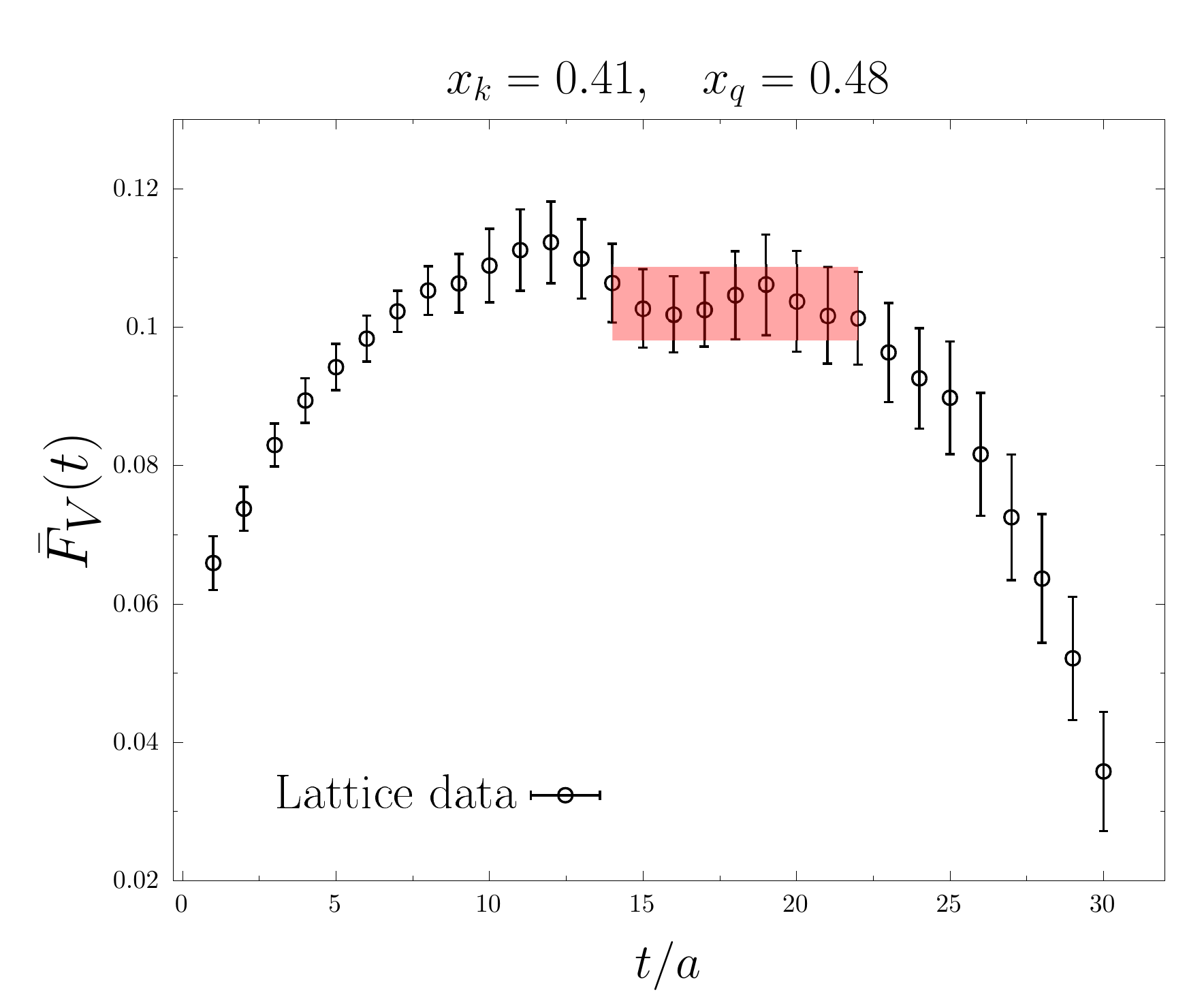}} \\
	\subfloat{%
		\includegraphics[scale=0.30]{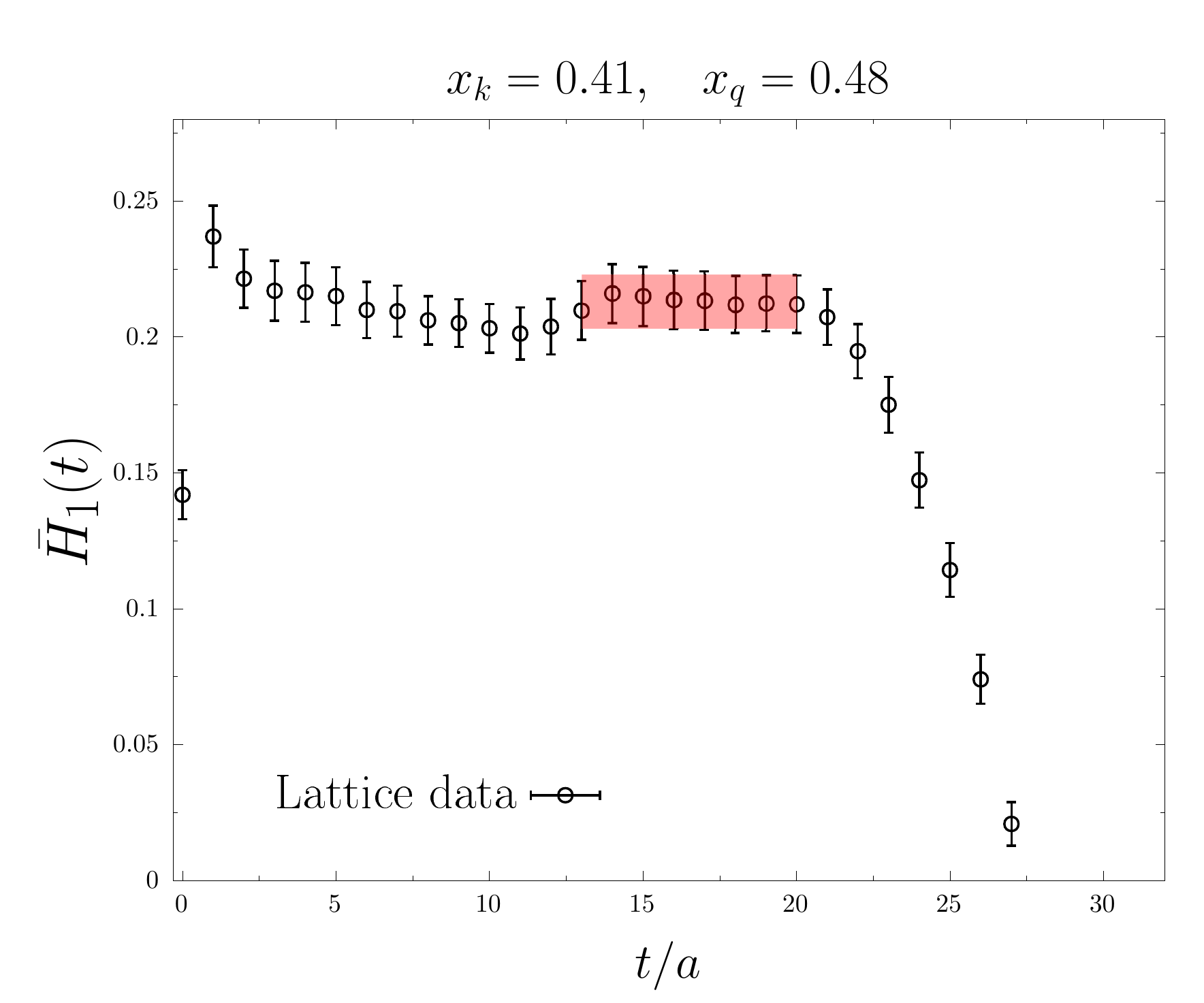}}
	\subfloat{%
		\includegraphics[scale=0.30]{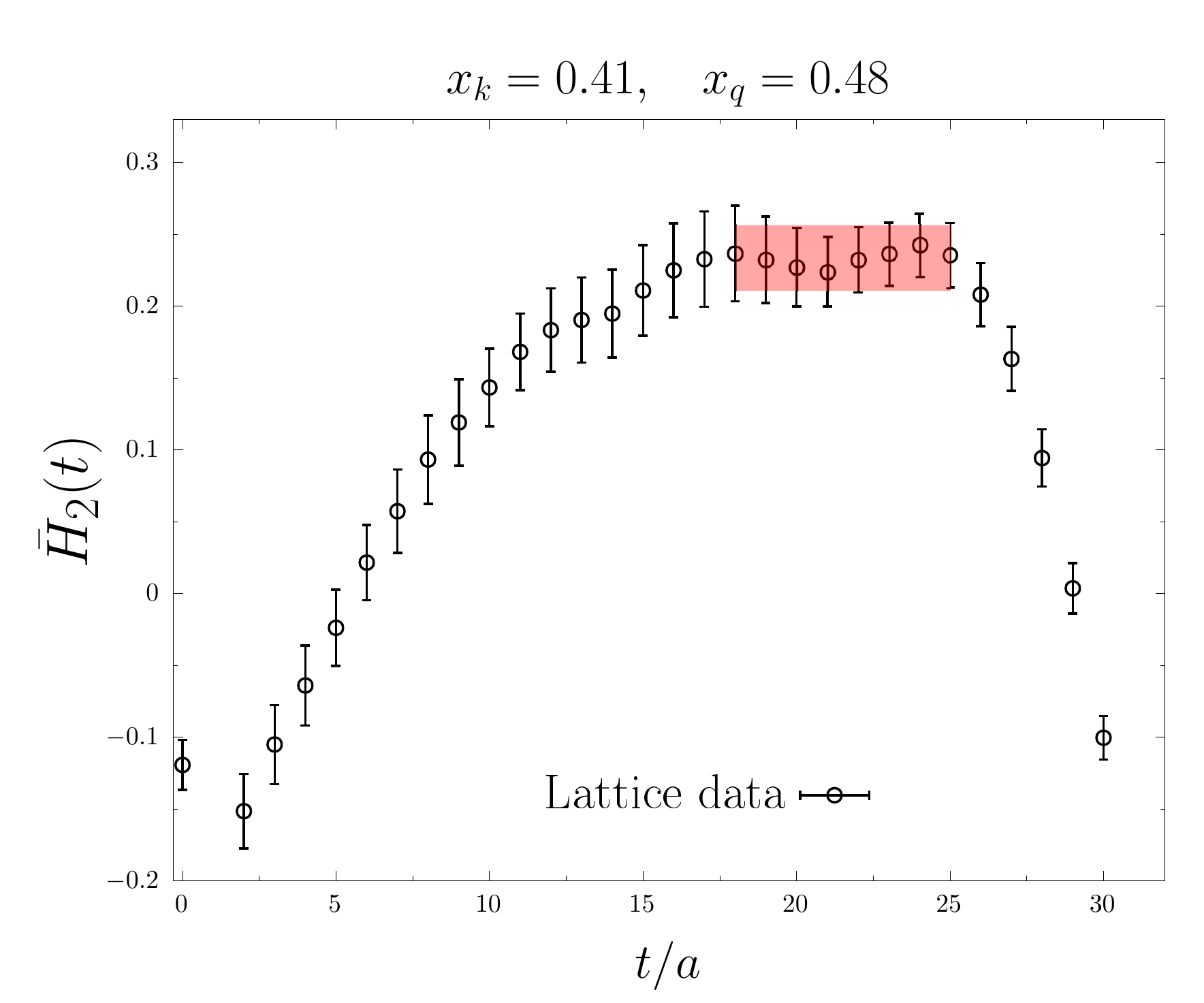}} 
	\caption{\it Extraction of the form factors $F_{A}, F_{V},H_{1}, H_{2}$ from the plateaux of the corresponding estimator. The data refer to the kinematic configuration with $x_{k}=0.41$ and $x_{q}=0.48$.}
	\label{xk_0.4064_xq_0.4845}
	\end{center}
\end{figure}

\begin{figure}
	\begin{center}
	\captionsetup{justification=justified,singlelinecheck=off}
	\subfloat{%
		\includegraphics[scale=0.30]{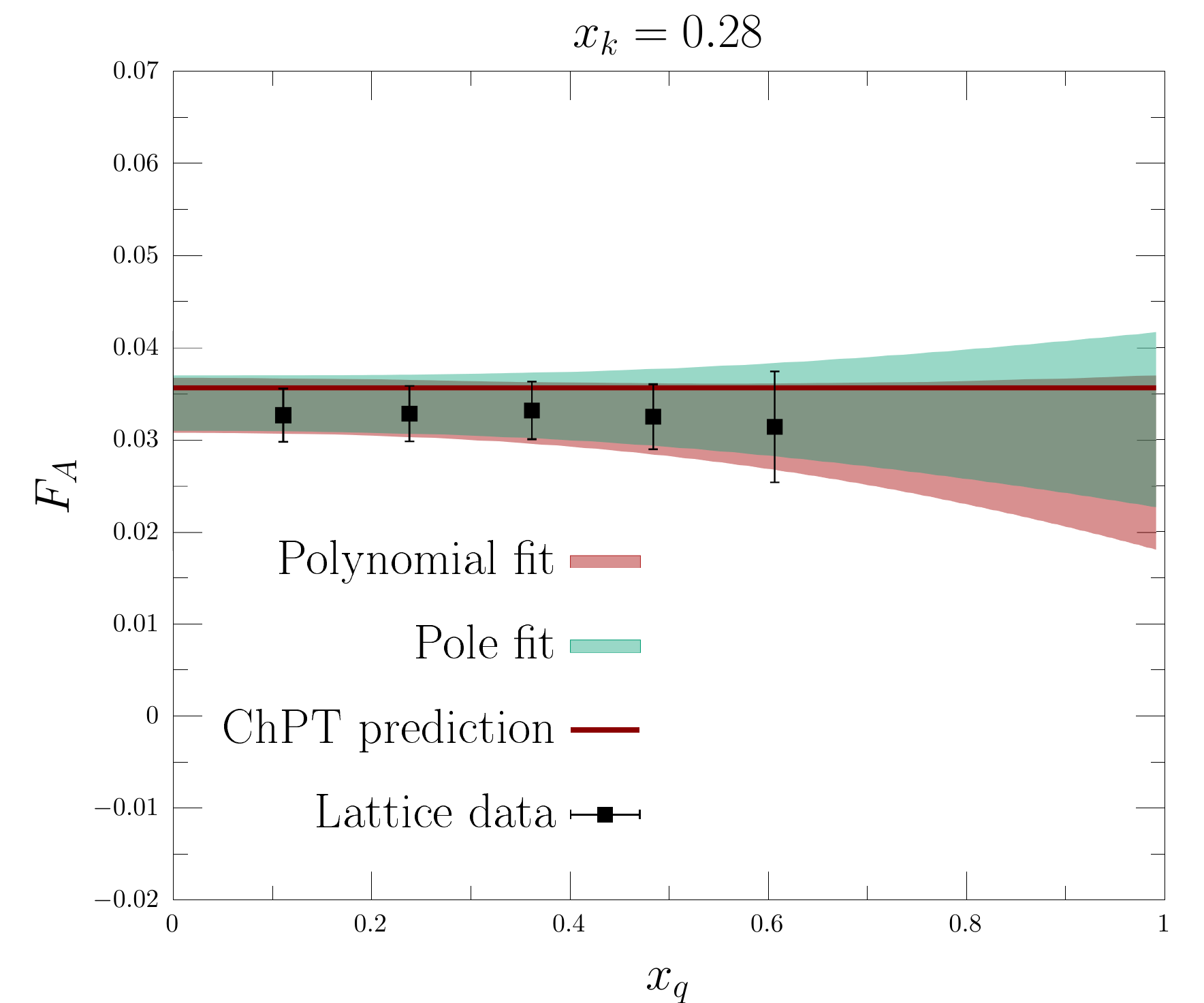}}
	\subfloat{%
		\includegraphics[scale=0.30]{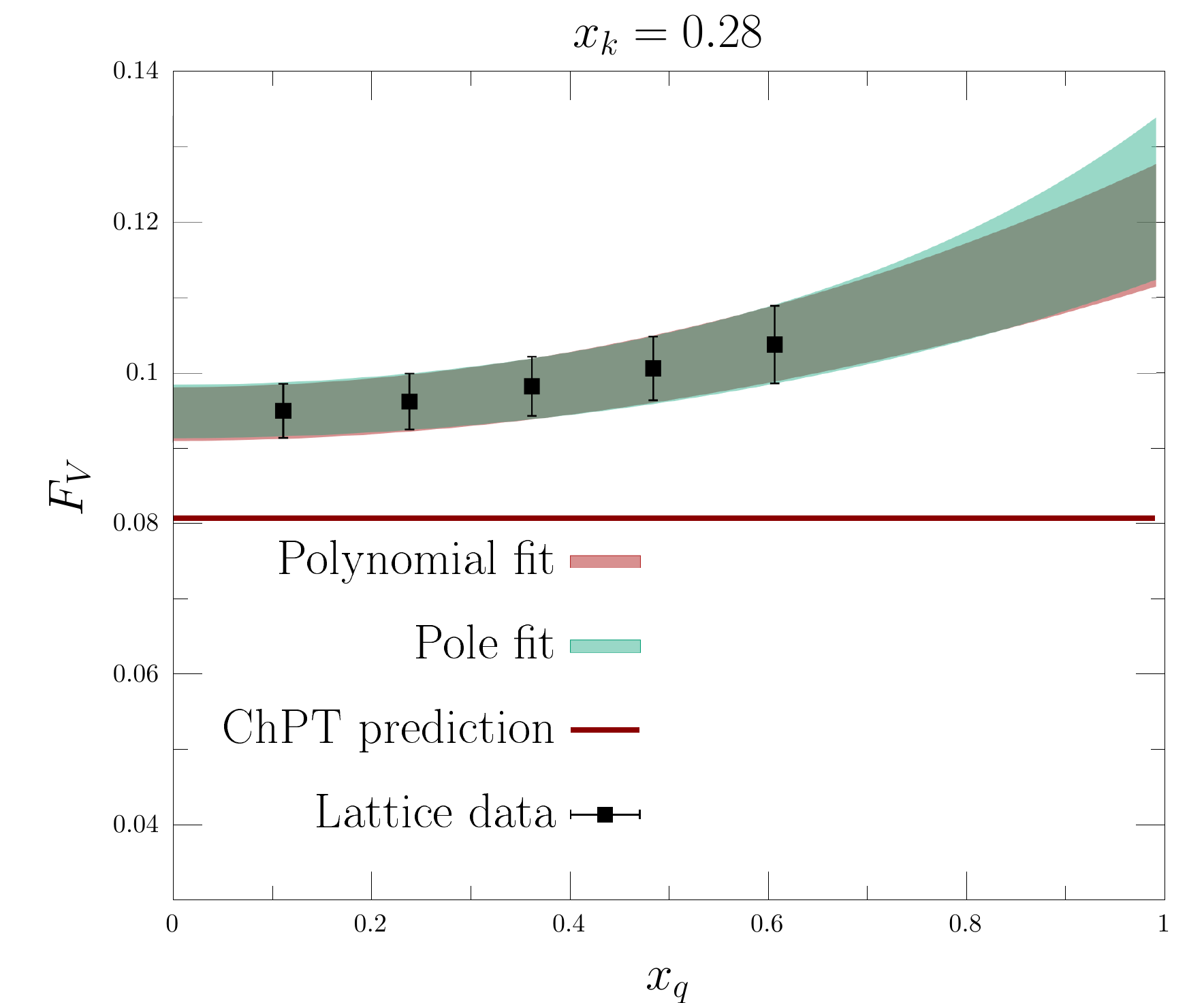}} \\
	\subfloat{%
		\includegraphics[scale=0.30]{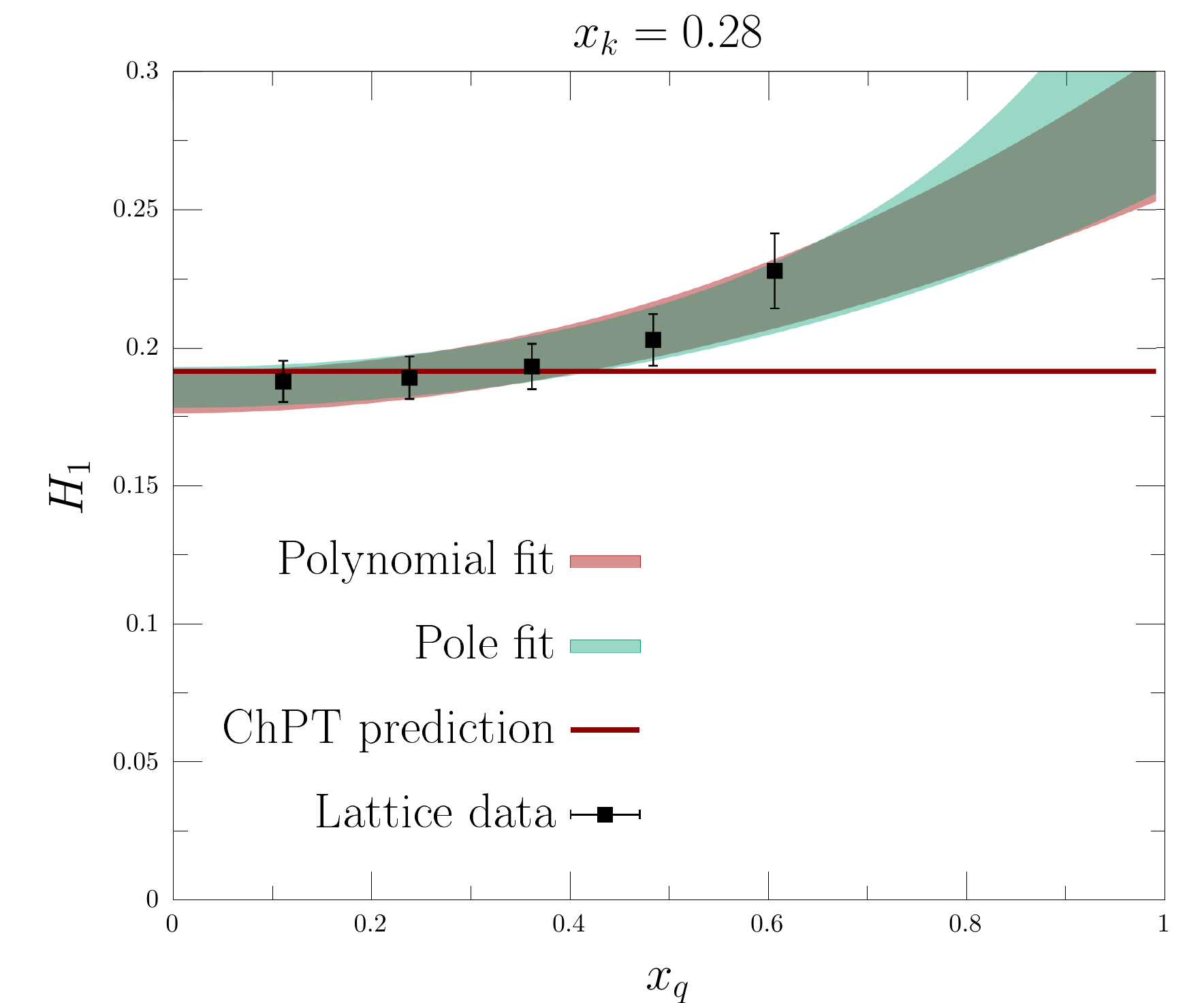}}
	\subfloat{%
		\includegraphics[scale=0.30]{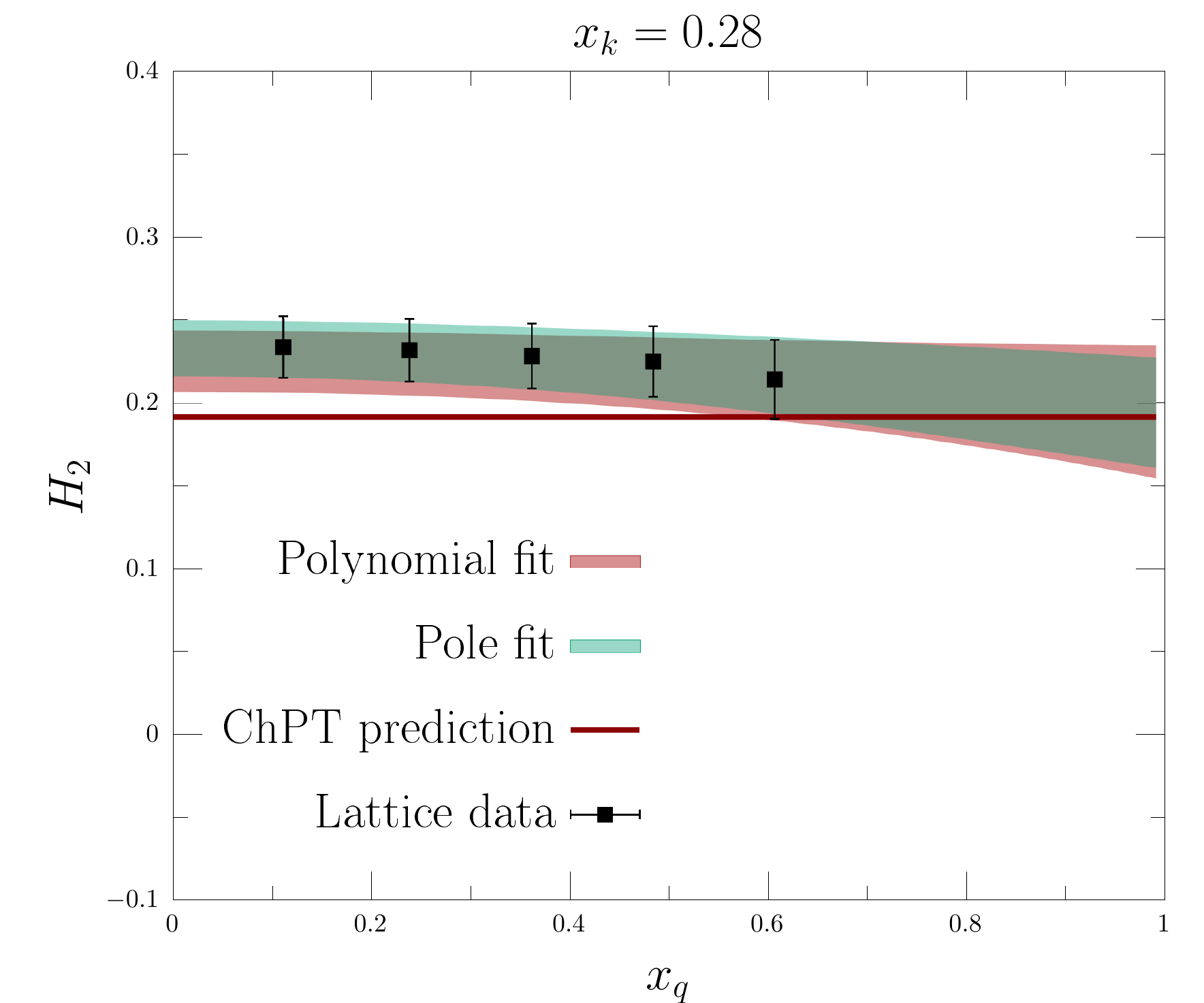}} 
	\caption{\it{The fitting functions corresponding to the polynomial and the pole-like fit of Eqs.~(\ref{poly}),~(\ref{pole}) are plotted, along with the lattice data, as function of $x_{q}$ and at a fixed value of $x_k=0.28$. The red line corresponds to the ChPT prediction at NLO.}\hspace*{\fill}
	\label{xk_0.2845}}
	\end{center}
\end{figure}

As already stated above, we can study separately the vector and axial components of the weak current. Since there is only one vector form factor, $F_V$, the vector component of the matrix element that we isolate from the lattice correlator is simply proportional to it and so we can extract it easily. 
For the axial form factors we proceed by defining three independent linear combinations of the form factors, where the point-like contribution proportional to $f_P$ has been subtracted. Then by inverting the coefficient matrix we are able to single out the three axial form factors $F_{A}$, $H_{1}$ and $H_{2}$.  
We now proceed to present our numerical results.
The simulations have been performed on the $A40.32$ ensemble generated by the ETMC~\cite{Carrasco2014} with $N_{f}=2+1+1$ dynamical quark flavors, which has a spatial extent $L=32$, aspect ratio $T/L = 2$, and corresponds to a lattice spacing $a= 0.0885(36)~{\rm fm}$.  The analysis
was performed on 100 gauge configurations.  The ensemble corresponds to an higher-than-physical pion mass $m_{\pi}\simeq 320~{\rm MeV}$, and to a kaon mass $m_{K}\sim 530~{\rm MeV}$. 
We used twisted boundary conditions \cite{de_Divitiis_2004} in order to evaluate the hadronic tensor for a range of values of the photon's spatial momentum $\boldsymbol{k}$. To probe the region of the phase-space relevant for the four $K\to \ell\,\nu_\ell\,\ell^{\prime+}\ell^{\prime-}$ decay channels, with $\ell,\ell' = e,\mu$, we evaluated the Euclidean three-point functions $C^{\mu\nu}(t,E_{\gamma}, \boldsymbol{k}, \boldsymbol{p})$ for fifteen different values of $(E_{\gamma}, \boldsymbol{k})$, with $\boldsymbol{k}= (0,0,k_{z})$, and restricted our analysis to the kaon rest frame $\boldsymbol{p}= 0$. We find it convenient to parametrize the phase space in terms of the two dimensionless parameters $x_{k}$ and $x_{q}$, defined as
\begin{align}\label{eq:xkqdef}
	x_k \equiv \sqrt{\frac{k^{2}}{m_{K}^{2}}}\, , \qquad x_q \equiv \sqrt{\frac{q^{2}}{m_{K}^{2}}}~, 
\end{align}
which are nothing but the invariant masses of the lepton-antilepton and of the lepton-neutrino pairs, normalized over the kaon mass $m_{K}$. In term of the lepton masses $m_{\ell}, m_{\ell'}$, the physical ranges for the values of $x_{k}$ and $x_{q}$ are given by
\begin{align}
	\label{x_ranges}
	\frac{2m_{\ell'}}{m_K}~\leq~x_k~\leq~1-\frac{m_\ell}{m_K},\qquad
	\frac{m_\ell}{m_K}~\leq ~x_q~\leq~ 1-x_k\,.
\end{align}
The numerical values of $x_{k}$ and $x_{q}$ are reported in Tab.~\ref{tab:num_val_xk_xq}. 
\begin{center}
	\small\addtolength{\tabcolsep}{-2pt}
	\begin{table}[]
		\begin{tabular}{c | c c c  c c c c c c c c c c c c c}
			\hline
			kin & 1 & 2 & 3 & 4 & 5 & 6 & 7 & 8 & 9 & 10 & 11 & 12 & 13 & 14 & 15 \\
			\hline
			$ x_{k}$ & 0.28 & 0.28 & 0.28 & 0.28 & 0.28 & 0.41 & 0.41 & 0.41 & 0.41 & 0.53 & 0.53 & 0.53 & 0.65 & 0.65 & 0.77 \\
			\hline
			$ x_{q}$ & 0.12 & 0.24 & 0.36 & 0.48 & 0.61 & 0.12 & 0.24 & 0.36 & 0.48 & 0.12 & 0.24 & 0.36 & 0.12 & 0.24 & 0.12 \\
			\hline
		\end{tabular}
		\caption{\small\it List of the values of $x_{k}$ and $x_{q}$ corresponding to the fifteen simulated kinematic configurations.}
		\label{tab:num_val_xk_xq}
	\end{table}
\end{center}
In Fig.~\ref{xk_0.4064_xq_0.4845}, we show a collection of the renormalized estimators $\overline{H}_{1}(t, x_{k}, x_{q})$, $\overline{H}_{2}(t, x_{k}, x_{q})$, $\overline{F}_{A}(t, x_{k}, x_{q})$, $\overline{F}_{V}(t, x_{k}, x_{q})$ for a specific kinematic. In each figure, the shadowed region indicates the result of a constant fit in the corresponding time interval. The precision achieved is in most of the kinematics very good and typically of order of five to ten percent.
In order to evaluate the decay rate, we decided to fit the lattice form factors, employing two different fitting ansatzes to describe their dependence on $x_{k}$ and $x_{k}$. The first one is a simple second order polynomial in $x_{k}^{2}$ and $x_{q}^{2}$ given by
\begin{align}
	F_{\mathrm{poly}}(x_k,x_q)=a_0+a_kx_k^2+a_qx_q^2+a_{kq}x_k^2x_q^2~,\label{poly} 
\end{align}
where $a_{0}, a_{k}, a_{q}$ and $a_{kq}$ are free fitting parameters,
while the second fitting function has a pole-like structure of the form
\begin{align}
	F_{\mathrm{pole}}(x_k,x_q)=\frac{A}{\left[\left(1-R_k x_k^2\right)\left(1-R_q x_q^2\right)\right]}\label{pole}~, 
\end{align}
where $A, R_{k}$ and $R_{q}$ are free fitting parameters. The resulting fitting curves, along with the lattice data and with the ChPT prediction, are shown, for all four form factors $H_{1}, H_{2}, F_{A}$ and $F_{V}$, in the panels of Fig.~\ref{xk_0.2845} as a function of $x_{q}$ at fixed $x_{k}$.
The quality of the fit is in all cases very good, with the reduced $\chi^{2}$ being always smaller than one. The fit parameters corresponding to the polynomial and to the pole-like fit are collected in Tab.~\ref{fitpar}. As evident for the figure, our results are reasonably consistent with the ChPT prediction\footnote{The next to leading order ChPT prediction for the form factors can be found in \cite{Bijnens:1994me} and at this order they are almost constant with respect to the exchanged momenta.}.

\begin{table}[h]
	\begin{center}
	\small
	\begin{tabular}{|l|l|l|l|l|l|l|l|}
		\hline
		& $a_0$        & $a_k$        & $a_q$         & $a_{kq}$      &  $A$       & $R_k$        & $R_q$        \\ \hline
		$H_1$ & 0.1745(88) & 0.122(28)   & 0.121(30)   & -0.30(12)    &  0.1792(78) & 0.453(88) & 0.40(10)   \\ \hline
		$H_2$ & 0.198(21) & 0.347(88)  & -0.02(4)   & -0.1(3)     & 0.217(17)  & 0.87(12)    & -0.2(2) \\ \hline
		$F_A$ & 0.0319(47) & 0.02(4)  & -0.037(13) & 0.42(20)  & 0.0320(30) & 0.74(50)  & 0.0(3)  \\ \hline
		$F_V$ & 0.0911(41)  & 0.045(18) & 0.0282(56)  & -0.035(32)  & 0.0921(38) & 0.38(13)  & 0.233(49)  \\ \hline
	\end{tabular}
	\caption{\it Values of the fit parameters for all the form factors, as obtained from the polynomial and pole-like fit of Eqs.~(\ref{poly}) and (\ref{pole}).}
	\label{fitpar}
	\end{center}
\end{table}

\section{Branching Ratios Numerical Results}
From the knowledge of the hadronic tensor $H_W^{\mu\nu}$, the $K^{+} \to \ell^{+} \nu_{l} \ell'^{+} \ell'^{-}$ decay rate is obtained integrating over the phase space of the final leptons and neutrino, the unpolarized squared amplitude $\sum_{\mathrm{spins}}|\mathcal{M}|^{2}$, where $\mathcal{M}$ is given by 
\begin{align}
	\label{l_not_lprime_amplitude}
	\mathcal{M}(p_{\ell'^{+}}, p_{\ell'^{-}}, p_{\ell^{+}}, p_{\nu_{\ell}})  =-\frac{G_F}{\sqrt{2}}V^*_{us}\frac{e^2}{k^2}\overline{u}(p_{\ell'^-})\gamma_\mu v(p_{\ell'^+})\left[f_{K}L^\mu(p_{\ell'^{+}}, p_{\ell'^{-}}, p_{\ell^{+}}, p_{\nu_{\ell}})-H_{\mathrm{SD}}^{\mu\nu}(p,q)l_{\nu}(p_{\ell^{+}}, p_{\nu_{\ell}})\right]~,    
\end{align}
where
\begin{align}
	L^{\mu}(p_{\ell'^{+}}, p_{\ell'^{-}}, p_{\ell^{+}}, p_{\nu_{\ell}}) &= m_\ell\overline{u}(p_{\nu_{\ell}})(1+\gamma_5)\left\{\frac{2p^\mu-k^\mu}{2pk-k^2}-\frac{2p_{\ell^{+}}^\mu+\slashed{k}\gamma^\mu}{2p_{\ell^{+}} k+k^2}\right\}v(p_{\ell^{+}})~,\\[8pt]
	l^{\mu}(p_{\ell^{+}}, p_{\nu_{\ell}}) &= \overline{u}(p_{\nu_{\ell}})\gamma^\mu(1-\gamma_5)v(p_{\ell^{+}})~.
\end{align}
In the previous equation, $p$ is the four-momentum of the kaon, $k=p_{\ell'^{+}}+p_{\ell'^{-}}$, and $q= p_{\ell^{+}}+ p_{\nu_{\ell}}$. In Eq.\,(\ref{l_not_lprime_amplitude}), the first term in the square parentheses gives the decay rate in the approximation in which the decaying kaon is treated as a point-like particle and includes the radiation from both the meson and charged lepton\,\footnote{This term is frequently referred to as the \textit{inner-bremsstrahlung} contribution.}. Except for the 
kaon decay constant $f_{K}$, the non-perturbative contribution to the rate is entirely contained in the second term of Eq.\,(\ref{l_not_lprime_amplitude}). When $\ell=\ell'$, since the final state leptons are indistinguishable, the exchange contribution, in which the momenta $p_{\ell'^{+}}$ and $p_{\ell^{+}}$ are interchanged, must be added to the amplitude $\mathcal{M}$ via the replacement
\begin{align}
	\label{l_ugual_lprime_amplitude}
	\mathcal{M}(p_{\ell'^{+}}, p_{\ell'^{-}}, p_{\ell^{+}}, p_{\nu_{\ell}}) \to \mathcal{M}(p_{\ell'^{+}}, p_{\ell'^{-}}, p_{\ell^{+}}, p_{\nu_{\ell}}) - \mathcal{M}(p_{\ell^{+}}, p_{\ell'^{-}}, p_{\ell'^{+}}, p_{\nu_{\ell}})~.   
\end{align}

In terms of $\sum_{\mathrm{spins}}|\mathcal{M}|^{2}$, the branching ratio for $K^{+}\to \ell^{+} \nu_{l} \ell'^{+} \ell'^{-}$ is given by 
\small
\bea
	& &\textrm{BR}\left[K^+\to \ell^{+} \nu_{l} \ell'^{+} \ell'^{-}\right] =\nonumber\\
	& &\frac{\mathcal{S}}{2m_K\Gamma_K (2\pi)^{8}}\int \sum_{spins}|\mathcal{M}|^2~\delta\left(p-p_{\ell^{+}}-p_{\nu_{\ell}}-p_{\ell'^+}-p_{\ell'^-}\right)\frac{d^3p_{\ell^{+}}}{2E_{\ell^{+}}}\frac{d^3p_{\nu_{\ell}}}{2E_{\nu_{\ell}}}\frac{d^3p_{\ell'^+}}{2E_{\ell'^+}}\frac{d^3p_{\ell'^-}}{2E_{\ell'^-}}~,
\eea
\normalsize
where $\Gamma_{K}= 5.3167(86)\times 10^{-17}~{\rm GeV}$ is the total decay rate of the $K^{+}$~\cite{pdg}, and $\mathcal{S}$ is a symmetry factor that takes the value $\mathcal{S}=1$ for $\ell \ne \ell'$ and $\mathcal{S}= 1/2$ for $\ell = \ell'$. 
We used the FeynCalc package of Mathematica~\cite{2020FenyCalc} to compute $\sum_{\mathrm{spins}}|\mathcal{M}|^{2}$. For the phase space integration we performed a Monte Carlo integration by employing the GSL implementation of the VEGAS algorithm  \cite{1978JCoPh..27..192L}.

\begin{figure}[h]
	\begin{center}
	\subfloat{%
		\includegraphics[scale=0.30]{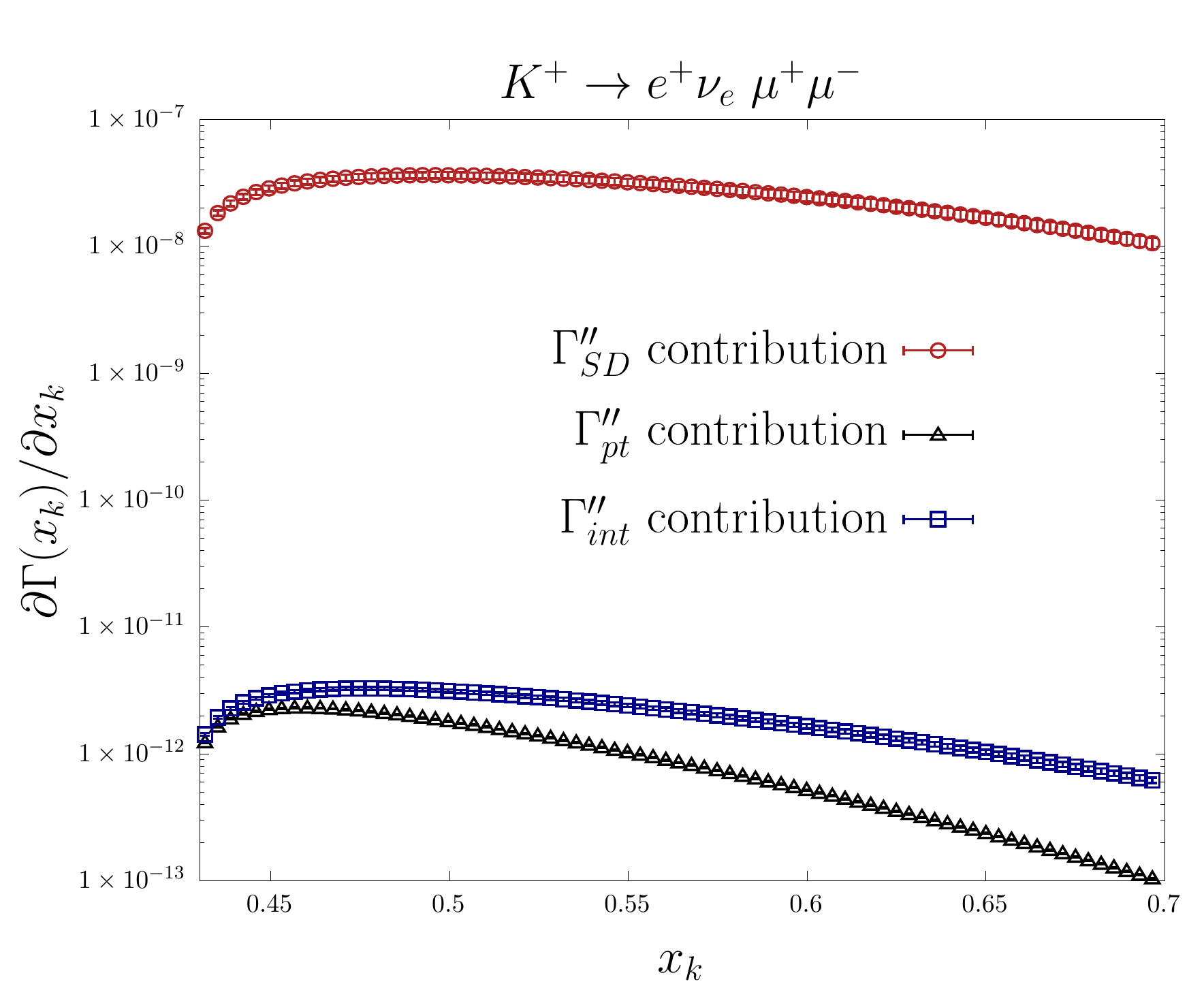}}
	\subfloat{%
		\includegraphics[scale=0.30]{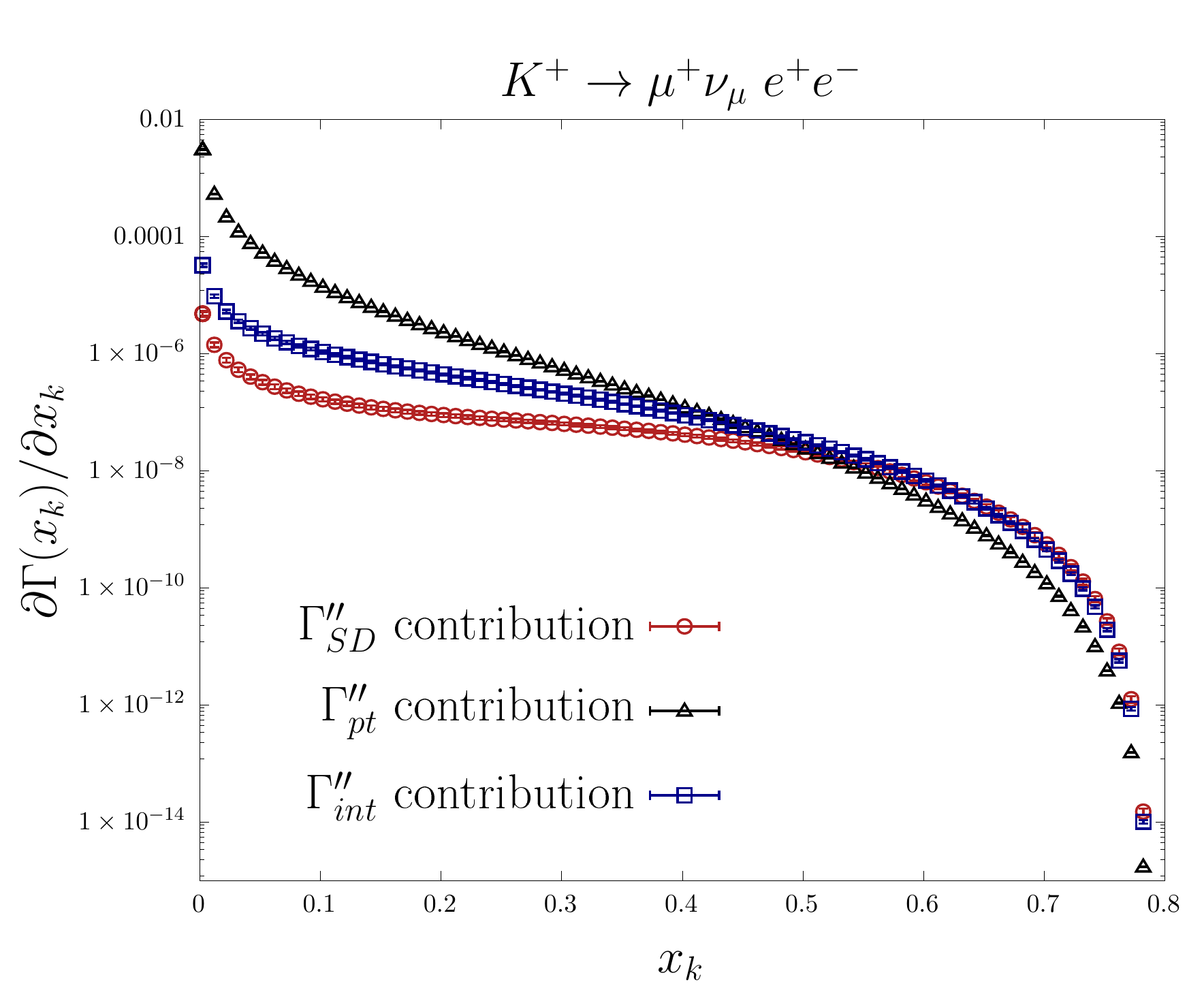}} \\
	\subfloat{%
		\includegraphics[scale=0.30]{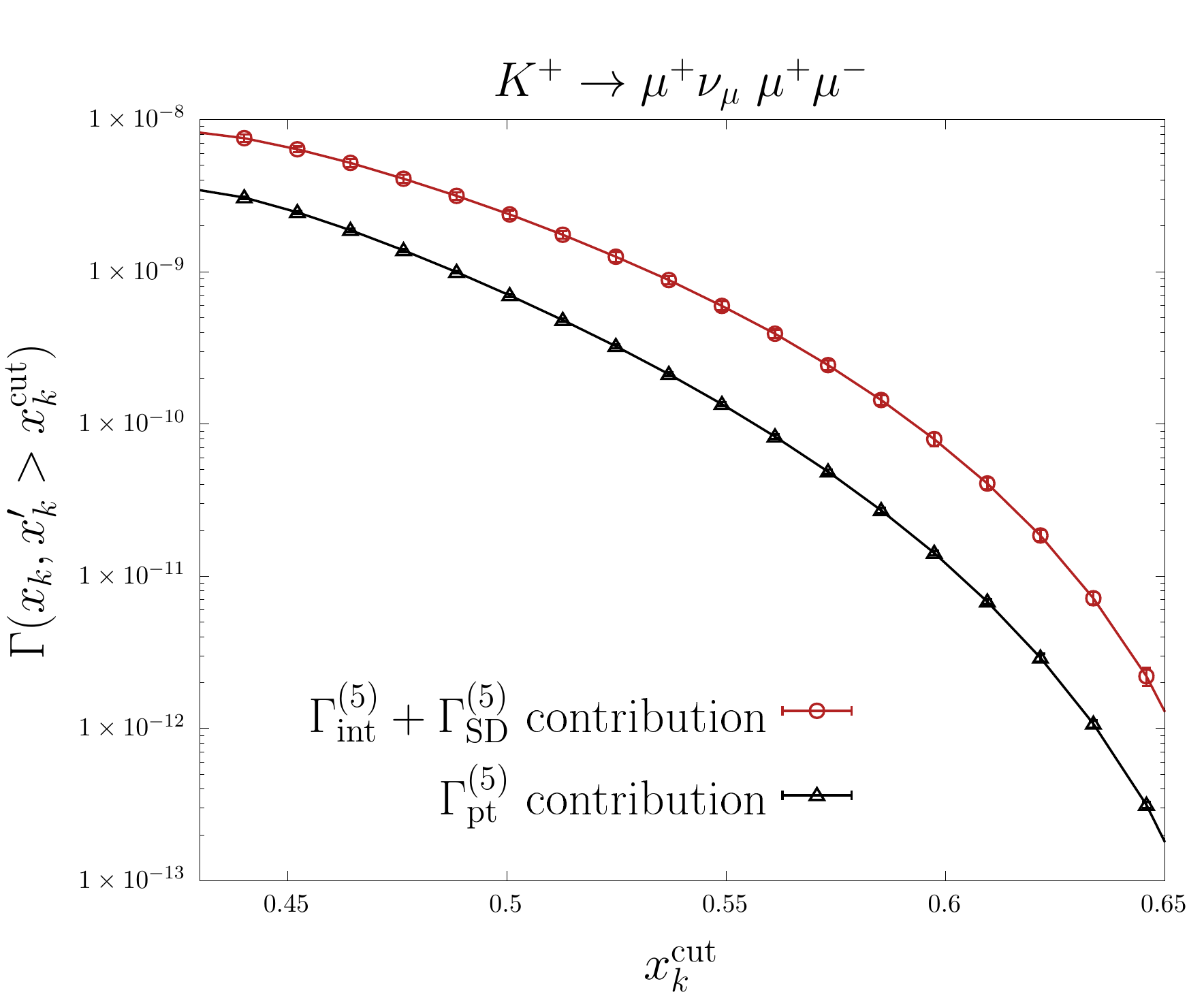}}
	\subfloat{%
		\includegraphics[scale=0.30]{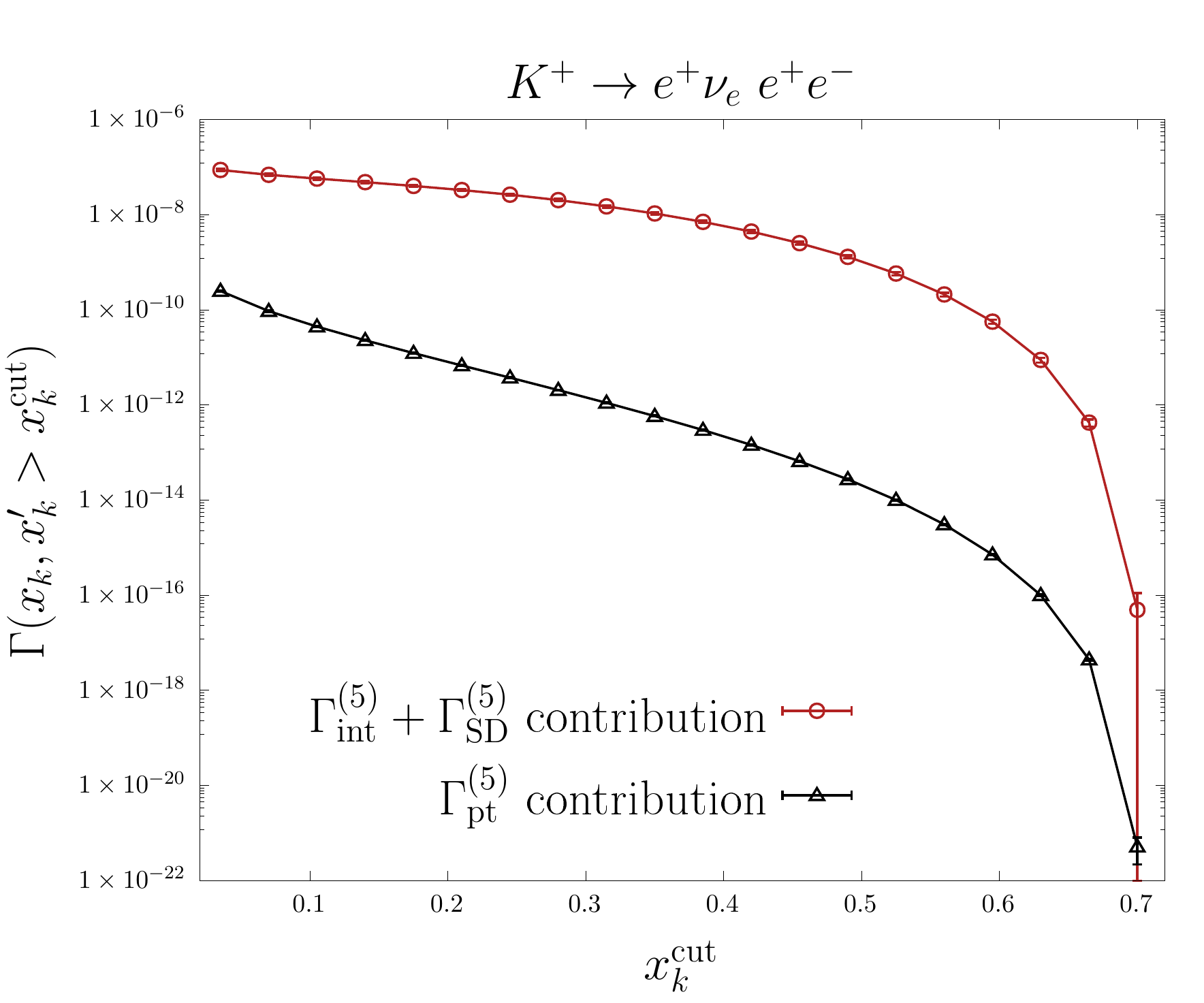}} 
	\caption{\it A comparison among the point-like, SD and interference contribution to the decay rate for the various channel. For the channels with different leptons in the final state we plot the differential decay rate $\partial \Gamma(x_k)/\partial x_k$. Instead, for the channels with same leptons in the final state, we plot the integrated decay rate $\Gamma(x_k,\,x_{k'}> x_{k}^{cut})$ as a function of the common lower cut on the invariant mass of the two possible $l^+\,l^-$ pairs.}
	\label{brplot}
	\end{center}
\end{figure}

Our final lattice predictions for the branching ratios are collected in Tab.~\ref{brres}, where we compare our result with the recent lattice determination of \cite{xu}, with the ChPT prediction\footnote{We reconstruct the branching ratios using the NLO ChPT formulae presented in \cite{Bijnens:1994me} for the SD form factors and setting the low energy constant $F=f_K/\sqrt{2}$.} and with the available experimental measurements.
We remark that a proper study of the systematic errors in both our numerical analysis and the one from \cite{xu} has yet to be done. In both cases, indeed, it has been employed only one gauge ensemble at unphysical lattice meson masses.
Finally, in Fig.~\ref{brplot} we show a comparison between the contribution to the decay rate coming from the point-like term, from the SD term and from the interference between the two. Clearly, all the information about the kaon internal structure (i.e. from $H^{\mu\nu}_{\mathrm{SD}}$) are contained in the latter two contributions. For the processes in which the weak current creates an electron, the SD contribution is by far the most dominant one in the branching ratios, because of the helicity suppression of the point-like term.
\begin{table*}
	\small
	\captionsetup{justification=justified,singlelinecheck=off}
	\begin{tabular}{ccccc}
		Channels & our Lattice & Lattice \cite{xu} & ChPT  & experiments\\
		\hline 
		$\operatorname{Br}[K\to \mu\nu_\mu e^+ e^-]$&$\quad8.28(14)\times 10^{-8}$ &  $\quad11.08(39)\times 10^{-8}$  &$\quad8.25\times 10^{-8}$&$\quad7.93(33)\times 10^{-8}$\cite{Poblaguev_2002}\\
		$\operatorname{Br}[K\to e\nu_e \mu^+ \mu^-]$&$\quad0.761(50)\times 10^{-8}$&$\quad0.94(8)\times 10^{-8}$ &$\quad0.62\times 10^{-8}$&$\quad1.72(45)\times 10^{-8}$\cite{Ma_2006}\\
		$\operatorname{Br}[K\to e\nu_e e^+ e^-]$ &$\quad1.95(11)\times 10^{-8}$ &$\quad3.29(35)\times 10^{-8}$ &$\quad1.75\times 10^{-8}$&$\quad2.91(23)\times 10^{-8}$\cite{Poblaguev_2002}\\
		$\operatorname{Br}[K\to \mu\nu_\mu \mu^+ \mu^-]$&$\quad1.178(37)\times 10^{-8}$  &$\quad1.52(7)\times 10^{-8}$&$\quad1.1.10\times 10^{-8}$&——\\
	\end{tabular}
	\caption{\it Comparison of our results for the branching ratios $\mathrm{Br}\left[K^+\to \ell^+\,\nu_{\ell}\,\ell'^+\,\ell'^-\right]$ with the results from Ref.\,\cite{xu}, the results obtained using the NLO ChPT predictions for the SD form factors and with the experiments \cite{Ma_2006, Poblaguev_2002}. The experimental results of $K\to e\nu_e e^+e^-$ and $K\to \mu\nu_\mu e^+e^-$ are the extrapolated values from $m_{ee}>150$ MeV and 145 MeV to $m_{ee}>140$ MeV, that is the lower cut we considered in our phase space integration for $m_{ee}$. The extrapolation formulae are given in Ref.~\cite{Poblaguev_2002}. We stress that our lattice determinations, as well as the ones from \cite{xu}, are affected by systematic uncertainties due to the missing continuum and physical point extrapolations.}
	\label{brres}
\end{table*}
\clearpage
\captionsetup{justification=justified,singlelinecheck=off}
\section{Conclusion}
We presented a strategy to compute, using Lattice QCD, the amplitudes and branching ratios for the decays $P\to\ell\nu_\ell\,\ell^{\prime\,+}\ell^{\prime\,-}$, where $P$ is a pseudoscalar meson and $\ell$ and $\ell^\prime$ are charged leptons. In particular, we demonstrate how the four structure-dependent (SD) form factors can be determined and separated from the point-like contribution.
We apply the developed formalism to the four channels of $K\to\ell\nu_\ell\,\ell^{\prime\,+}\ell^{\prime\,-}$ decays, where $\ell$ and $\ell^\prime=\mu$ or $e$, in an exploratory Lattice QCD computation at a single lattice spacing and at unphysical light-quark masses. We demonstrate that all four SD form factors, $F_V,\,F_A,\,H_1$ and $H_2$ can be determined with good precision and used to calculate the corresponding branching ratios. In spite of the unphysical quark masses used in this simulation, it has been interesting and instructive to compare our results with those from experiment (where available) and from NLO ChPT. As can be seen from Tab.\,\ref{brres},
the results are generally in reasonable semi-quantitative agreement.
Our future work will focus on controlling and reducing the systematic uncertainties resulting from the absence of continuum, chiral and infinite-volume extrapolations. 
We will also work to extend the method to heavier pseudoscalar mesons, for which the analytic continuation to Euclidean time gives rise to enhanced finite-volume effects due to the presence of internal lighter states. A theoretical, model independent, prediction for such processes will be very useful to test the validity of the Standard Model and for the search of New Physics.
\section{Acknowledgment}
We thank C. Tarantino for useful discussions, and all members of the ETMC for the most enjoyable collaboration. We acknowledge CINECA for the provision of CPU time under the specific initiative INFN-LQCD123 and IscrB\_S-EPIC. F.S. G.G and S.S. are supported by the Italian Ministry of University and Research (MIUR) under grant PRIN20172LNEEZ. F.S. and G.G are supported by INFN under GRANT73/CALAT. C.T.S. was partially supported by an Emeritus Fellowship from the Leverhulme Trust and by STFC (UK) grants ST/P000711/1 and ST/T000775/1.

\bibliographystyle{apsrev4-2}
\bibliography{bibliography}

\begin{thebibliography}{14}%
\makeatletter
\providecommand \@ifxundefined [1]{%
 \@ifx{#1\undefined}
}%
\providecommand \@ifnum [1]{%
 \ifnum #1\expandafter \@firstoftwo
 \else \expandafter \@secondoftwo
 \fi
}%
\providecommand \@ifx [1]{%
 \ifx #1\expandafter \@firstoftwo
 \else \expandafter \@secondoftwo
 \fi
}%
\providecommand \natexlab [1]{#1}%
\providecommand \enquote  [1]{``#1''}%
\providecommand \bibnamefont  [1]{#1}%
\providecommand \bibfnamefont [1]{#1}%
\providecommand \citenamefont [1]{#1}%
\providecommand \href@noop [0]{\@secondoftwo}%
\providecommand \href [0]{\begingroup \@sanitize@url \@href}%
\providecommand \@href[1]{\@@startlink{#1}\@@href}%
\providecommand \@@href[1]{\endgroup#1\@@endlink}%
\providecommand \@sanitize@url [0]{\catcode `\\12\catcode `\$12\catcode
  `\&12\catcode `\#12\catcode `\^12\catcode `\_12\catcode `\%12\relax}%
\providecommand \@@startlink[1]{}%
\providecommand \@@endlink[0]{}%
\providecommand \url  [0]{\begingroup\@sanitize@url \@url }%
\providecommand \@url [1]{\endgroup\@href {#1}{\urlprefix }}%
\providecommand \urlprefix  [0]{URL }%
\providecommand \Eprint [0]{\href }%
\providecommand \doibase [0]{https://doi.org/}%
\providecommand \selectlanguage [0]{\@gobble}%
\providecommand \bibinfo  [0]{\@secondoftwo}%
\providecommand \bibfield  [0]{\@secondoftwo}%
\providecommand \translation [1]{[#1]}%
\providecommand \BibitemOpen [0]{}%
\providecommand \bibitemStop [0]{}%
\providecommand \bibitemNoStop [0]{.\EOS\space}%
\providecommand \EOS [0]{\spacefactor3000\relax}%
\providecommand \BibitemShut  [1]{\csname bibitem#1\endcsname}%
\let\auto@bib@innerbib\@empty
%</preamble>
\bibitem [{\citenamefont {Bijnens}\ \emph {et~al.}(1994)\citenamefont
  {Bijnens}, \citenamefont {Colangelo}, \citenamefont {Ecker},\ and\
  \citenamefont {Gasser}}]{Bijnens:1994me}%
  \BibitemOpen
  \bibfield  {author} {\bibinfo {author} {\bibfnamefont {J.}~\bibnamefont
  {Bijnens}}, \bibinfo {author} {\bibfnamefont {G.}~\bibnamefont {Colangelo}},
  \bibinfo {author} {\bibfnamefont {G.}~\bibnamefont {Ecker}},\ and\ \bibinfo
  {author} {\bibfnamefont {J.}~\bibnamefont {Gasser}}\ }(\bibinfo {year}
  {1994})\ \Eprint {https://arxiv.org/abs/hep-ph/9411311}
  {arXiv:hep-ph/9411311} \BibitemShut {NoStop}%
\bibitem [{\citenamefont {Danilina}\ \emph {et~al.}(2020)\citenamefont
  {Danilina}, \citenamefont {Nikitin},\ and\ \citenamefont
  {Toms}}]{Danilina_2020}%
  \BibitemOpen
  \bibfield  {author} {\bibinfo {author} {\bibfnamefont {A.}~\bibnamefont
  {Danilina}}, \bibinfo {author} {\bibfnamefont {N.}~\bibnamefont {Nikitin}},\
  and\ \bibinfo {author} {\bibfnamefont {K.}~\bibnamefont {Toms}},\ }\bibfield
  {journal} {\bibinfo  {journal} {Physical Review D}\ }\textbf {\bibinfo
  {volume} {101}},\ \href {https://doi.org/10.1103/physrevd.101.096007}
  {10.1103/physrevd.101.096007} (\bibinfo {year} {2020})\BibitemShut {NoStop}%
\bibitem [{\citenamefont {Aaij}\ \emph {et~al.}(2019)\citenamefont {Aaij},
  \citenamefont {Beteta}, \citenamefont {Adeva}, \citenamefont {Adinolfi},
  \citenamefont {Aidala}, \citenamefont {Ajaltouni}, \citenamefont {Akar},
  \citenamefont {Albicocco}, \citenamefont {Albrecht},\ and\ \citenamefont
  {et~al.}}]{Aaij_2019}%
  \BibitemOpen
  \bibfield  {author} {\bibinfo {author} {\bibfnamefont {R.}~\bibnamefont
  {Aaij}}, \bibinfo {author} {\bibfnamefont {C.~A.}\ \bibnamefont {Beteta}},
  \bibinfo {author} {\bibfnamefont {B.}~\bibnamefont {Adeva}}, \bibinfo
  {author} {\bibfnamefont {M.}~\bibnamefont {Adinolfi}}, \bibinfo {author}
  {\bibfnamefont {C.~A.}\ \bibnamefont {Aidala}}, \bibinfo {author}
  {\bibfnamefont {Z.}~\bibnamefont {Ajaltouni}}, \bibinfo {author}
  {\bibfnamefont {S.}~\bibnamefont {Akar}}, \bibinfo {author} {\bibfnamefont
  {P.}~\bibnamefont {Albicocco}}, \bibinfo {author} {\bibfnamefont
  {J.}~\bibnamefont {Albrecht}},\ and\ \bibinfo {author} {\bibnamefont
  {et~al.}},\ }\bibfield  {journal} {\bibinfo  {journal} {The European Physical
  Journal C}\ }\textbf {\bibinfo {volume} {79}},\ \href
  {https://doi.org/10.1140/epjc/s10052-019-7112-x}
  {10.1140/epjc/s10052-019-7112-x} (\bibinfo {year} {2019})\BibitemShut
  {NoStop}%
\bibitem [{\citenamefont {Tuo}\ \emph {et~al.}(2021)\citenamefont {Tuo},
  \citenamefont {Feng}, \citenamefont {Jin},\ and\ \citenamefont {Wang}}]{xu}%
  \BibitemOpen
  \bibfield  {author} {\bibinfo {author} {\bibfnamefont {X.-Y.}\ \bibnamefont
  {Tuo}}, \bibinfo {author} {\bibfnamefont {X.}~\bibnamefont {Feng}}, \bibinfo
  {author} {\bibfnamefont {L.-C.}\ \bibnamefont {Jin}},\ and\ \bibinfo {author}
  {\bibfnamefont {T.}~\bibnamefont {Wang}},\ }\href@noop {} {} (\bibinfo {year}
  {2021}),\ \Eprint {https://arxiv.org/abs/2103.11331} {arXiv:2103.11331
  [hep-lat]} \BibitemShut {NoStop}%
\bibitem [{\citenamefont {Carrasco}\ \emph {et~al.}(2015)\citenamefont
  {Carrasco}, \citenamefont {Lubicz}, \citenamefont {Martinelli}, \citenamefont
  {Sachrajda}, \citenamefont {Tantalo}, \citenamefont {Tarantino},\ and\
  \citenamefont {Testa}}]{Carrasco_2015}%
  \BibitemOpen
  \bibfield  {author} {\bibinfo {author} {\bibfnamefont {N.}~\bibnamefont
  {Carrasco}}, \bibinfo {author} {\bibfnamefont {V.}~\bibnamefont {Lubicz}},
  \bibinfo {author} {\bibfnamefont {G.}~\bibnamefont {Martinelli}}, \bibinfo
  {author} {\bibfnamefont {C.}~\bibnamefont {Sachrajda}}, \bibinfo {author}
  {\bibfnamefont {N.}~\bibnamefont {Tantalo}}, \bibinfo {author} {\bibfnamefont
  {C.}~\bibnamefont {Tarantino}},\ and\ \bibinfo {author} {\bibfnamefont
  {M.}~\bibnamefont {Testa}},\ }\bibfield  {journal} {\bibinfo  {journal}
  {Physical Review D}\ }\textbf {\bibinfo {volume} {91}},\ \href
  {https://doi.org/10.1103/physrevd.91.074506} {10.1103/physrevd.91.074506}
  (\bibinfo {year} {2015})\BibitemShut {NoStop}%
\bibitem [{\citenamefont {Maiani}\ and\ \citenamefont {Testa}(1990)}]{maiani}%
  \BibitemOpen
  \bibfield  {author} {\bibinfo {author} {\bibfnamefont {L.}~\bibnamefont
  {Maiani}}\ and\ \bibinfo {author} {\bibfnamefont {M.}~\bibnamefont {Testa}},\
  }\href {https://doi.org/https://doi.org/10.1016/0370-2693(90)90695-3}
  {\bibfield  {journal} {\bibinfo  {journal} {Physics Letters B}\ }\textbf
  {\bibinfo {volume} {245}},\ \bibinfo {pages} {585} (\bibinfo {year}
  {1990})}\BibitemShut {NoStop}%
\bibitem [{\citenamefont {Desiderio}\ \emph {et~al.}(2021)\citenamefont
  {Desiderio}, \citenamefont {Frezzotti}, \citenamefont {Garofalo},
  \citenamefont {Giusti}, \citenamefont {Hansen}, \citenamefont {Lubicz},
  \citenamefont {Martinelli}, \citenamefont {Sachrajda}, \citenamefont
  {Sanfilippo}, \citenamefont {Simula},\ and\ \citenamefont
  {et~al.}}]{Desiderio2020}%
  \BibitemOpen
  \bibfield  {author} {\bibinfo {author} {\bibfnamefont {A.}~\bibnamefont
  {Desiderio}}, \bibinfo {author} {\bibfnamefont {R.}~\bibnamefont
  {Frezzotti}}, \bibinfo {author} {\bibfnamefont {M.}~\bibnamefont {Garofalo}},
  \bibinfo {author} {\bibfnamefont {D.}~\bibnamefont {Giusti}}, \bibinfo
  {author} {\bibfnamefont {M.}~\bibnamefont {Hansen}}, \bibinfo {author}
  {\bibfnamefont {V.}~\bibnamefont {Lubicz}}, \bibinfo {author} {\bibfnamefont
  {G.}~\bibnamefont {Martinelli}}, \bibinfo {author} {\bibfnamefont
  {C.}~\bibnamefont {Sachrajda}}, \bibinfo {author} {\bibfnamefont
  {F.}~\bibnamefont {Sanfilippo}}, \bibinfo {author} {\bibfnamefont
  {S.}~\bibnamefont {Simula}},\ and\ \bibinfo {author} {\bibnamefont
  {et~al.}},\ }\bibfield  {journal} {\bibinfo  {journal} {Physical Review D}\
  }\textbf {\bibinfo {volume} {103}},\ \href
  {https://doi.org/10.1103/physrevd.103.014502} {10.1103/physrevd.103.014502}
  (\bibinfo {year} {2021})\BibitemShut {NoStop}%
\bibitem [{\citenamefont {Carrasco}\ \emph {et~al.}(2014)\citenamefont
  {Carrasco}, \citenamefont {Deuzeman}, \citenamefont {Dimopoulos},
  \citenamefont {Frezzotti}, \citenamefont {Giménez}, \citenamefont
  {Herdoiza}, \citenamefont {Lami}, \citenamefont {Lubicz}, \citenamefont
  {Palao}, \citenamefont {Picca},\ and\ \citenamefont {et~al.}}]{Carrasco2014}%
  \BibitemOpen
  \bibfield  {author} {\bibinfo {author} {\bibfnamefont {N.}~\bibnamefont
  {Carrasco}}, \bibinfo {author} {\bibfnamefont {A.}~\bibnamefont {Deuzeman}},
  \bibinfo {author} {\bibfnamefont {P.}~\bibnamefont {Dimopoulos}}, \bibinfo
  {author} {\bibfnamefont {R.}~\bibnamefont {Frezzotti}}, \bibinfo {author}
  {\bibfnamefont {V.}~\bibnamefont {Giménez}}, \bibinfo {author}
  {\bibfnamefont {G.}~\bibnamefont {Herdoiza}}, \bibinfo {author}
  {\bibfnamefont {P.}~\bibnamefont {Lami}}, \bibinfo {author} {\bibfnamefont
  {V.}~\bibnamefont {Lubicz}}, \bibinfo {author} {\bibfnamefont
  {D.}~\bibnamefont {Palao}}, \bibinfo {author} {\bibfnamefont
  {E.}~\bibnamefont {Picca}},\ and\ \bibinfo {author} {\bibnamefont {et~al.}},\
  }\href {https://doi.org/10.1016/j.nuclphysb.2014.07.025} {\bibfield
  {journal} {\bibinfo  {journal} {Nuclear Physics B}\ }\textbf {\bibinfo
  {volume} {887}},\ \bibinfo {pages} {19–68} (\bibinfo {year}
  {2014})}\BibitemShut {NoStop}%
\bibitem [{\citenamefont {de~Divitiis}\ \emph {et~al.}(2004)\citenamefont
  {de~Divitiis}, \citenamefont {Petronzio},\ and\ \citenamefont
  {Tantalo}}]{de_Divitiis_2004}%
  \BibitemOpen
  \bibfield  {author} {\bibinfo {author} {\bibfnamefont {G.}~\bibnamefont
  {de~Divitiis}}, \bibinfo {author} {\bibfnamefont {R.}~\bibnamefont
  {Petronzio}},\ and\ \bibinfo {author} {\bibfnamefont {N.}~\bibnamefont
  {Tantalo}},\ }\href {https://doi.org/10.1016/j.physletb.2004.06.035}
  {\bibfield  {journal} {\bibinfo  {journal} {Physics Letters B}\ }\textbf
  {\bibinfo {volume} {595}},\ \bibinfo {pages} {408–413} (\bibinfo {year}
  {2004})}\BibitemShut {NoStop}%
\bibitem [{\citenamefont {Zyla}\ \emph {et~al.}(2020)\citenamefont {Zyla} \emph
  {et~al.}}]{pdg}%
  \BibitemOpen
  \bibfield  {author} {\bibinfo {author} {\bibfnamefont {P.}~\bibnamefont
  {Zyla}} \emph {et~al.} (\bibinfo {collaboration} {Particle Data Group}),\
  }\href {https://doi.org/10.1093/ptep/ptaa104} {\bibfield  {journal} {\bibinfo
   {journal} {PTEP}\ }\textbf {\bibinfo {volume} {2020}},\ \bibinfo {pages}
  {083C01} (\bibinfo {year} {2020})}\BibitemShut {NoStop}%
\bibitem [{\citenamefont {Shtabovenko}\ \emph {et~al.}(2020)\citenamefont
  {Shtabovenko}, \citenamefont {Mertig},\ and\ \citenamefont
  {Orellana}}]{2020FenyCalc}%
  \BibitemOpen
  \bibfield  {author} {\bibinfo {author} {\bibfnamefont {V.}~\bibnamefont
  {Shtabovenko}}, \bibinfo {author} {\bibfnamefont {R.}~\bibnamefont
  {Mertig}},\ and\ \bibinfo {author} {\bibfnamefont {F.}~\bibnamefont
  {Orellana}},\ }\href {https://doi.org/10.1016/j.cpc.2020.107478} {\bibfield
  {journal} {\bibinfo  {journal} {Computer Physics Communications}\ }\textbf
  {\bibinfo {volume} {256}},\ \bibinfo {pages} {107478} (\bibinfo {year}
  {2020})}\BibitemShut {NoStop}%
\bibitem [{\citenamefont {{Lepage}}(1978)}]{1978JCoPh..27..192L}%
  \BibitemOpen
  \bibfield  {author} {\bibinfo {author} {\bibfnamefont {G.~P.}\ \bibnamefont
  {{Lepage}}},\ }\href {https://doi.org/10.1016/0021-9991(78)90004-9}
  {\bibfield  {journal} {\bibinfo  {journal} {Journal of Computational
  Physics}\ }\textbf {\bibinfo {volume} {27}},\ \bibinfo {pages} {192}
  (\bibinfo {year} {1978})}\BibitemShut {NoStop}%
\bibitem [{\citenamefont {Poblaguev}\ \emph {et~al.}(2002)\citenamefont
  {Poblaguev}, \citenamefont {Appel}, \citenamefont {Atoyan}, \citenamefont
  {Bassalleck}, \citenamefont {Bergman}, \citenamefont {Cheung}, \citenamefont
  {Dhawan}, \citenamefont {Do}, \citenamefont {Egger}, \citenamefont
  {Eilerts},\ and\ \citenamefont {et~al.}}]{Poblaguev_2002}%
  \BibitemOpen
  \bibfield  {author} {\bibinfo {author} {\bibfnamefont {A.~A.}\ \bibnamefont
  {Poblaguev}}, \bibinfo {author} {\bibfnamefont {R.}~\bibnamefont {Appel}},
  \bibinfo {author} {\bibfnamefont {G.~S.}\ \bibnamefont {Atoyan}}, \bibinfo
  {author} {\bibfnamefont {B.}~\bibnamefont {Bassalleck}}, \bibinfo {author}
  {\bibfnamefont {D.~R.}\ \bibnamefont {Bergman}}, \bibinfo {author}
  {\bibfnamefont {N.}~\bibnamefont {Cheung}}, \bibinfo {author} {\bibfnamefont
  {S.}~\bibnamefont {Dhawan}}, \bibinfo {author} {\bibfnamefont
  {H.}~\bibnamefont {Do}}, \bibinfo {author} {\bibfnamefont {J.}~\bibnamefont
  {Egger}}, \bibinfo {author} {\bibfnamefont {S.}~\bibnamefont {Eilerts}},\
  and\ \bibinfo {author} {\bibnamefont {et~al.}},\ }\bibfield  {journal}
  {\bibinfo  {journal} {Physical Review Letters}\ }\textbf {\bibinfo {volume}
  {89}},\ \href {https://doi.org/10.1103/physrevlett.89.061803}
  {10.1103/physrevlett.89.061803} (\bibinfo {year} {2002})\BibitemShut
  {NoStop}%
\bibitem [{\citenamefont {Ma}\ \emph {et~al.}(2006)\citenamefont {Ma},
  \citenamefont {Appel}, \citenamefont {Atoyan}, \citenamefont {Bassalleck},
  \citenamefont {Bergman}, \citenamefont {Cheung}, \citenamefont {Dhawan},
  \citenamefont {Do}, \citenamefont {Egger}, \citenamefont {Eilerts},\ and\
  \citenamefont {et~al.}}]{Ma_2006}%
  \BibitemOpen
  \bibfield  {author} {\bibinfo {author} {\bibfnamefont {H.}~\bibnamefont
  {Ma}}, \bibinfo {author} {\bibfnamefont {R.}~\bibnamefont {Appel}}, \bibinfo
  {author} {\bibfnamefont {G.~S.}\ \bibnamefont {Atoyan}}, \bibinfo {author}
  {\bibfnamefont {B.}~\bibnamefont {Bassalleck}}, \bibinfo {author}
  {\bibfnamefont {D.~R.}\ \bibnamefont {Bergman}}, \bibinfo {author}
  {\bibfnamefont {N.}~\bibnamefont {Cheung}}, \bibinfo {author} {\bibfnamefont
  {S.}~\bibnamefont {Dhawan}}, \bibinfo {author} {\bibfnamefont
  {H.}~\bibnamefont {Do}}, \bibinfo {author} {\bibfnamefont {J.}~\bibnamefont
  {Egger}}, \bibinfo {author} {\bibfnamefont {S.}~\bibnamefont {Eilerts}},\
  and\ \bibinfo {author} {\bibnamefont {et~al.}},\ }\bibfield  {journal}
  {\bibinfo  {journal} {Physical Review D}\ }\textbf {\bibinfo {volume} {73}},\
  \href {https://doi.org/10.1103/physrevd.73.037101}
  {10.1103/physrevd.73.037101} (\bibinfo {year} {2006})\BibitemShut {NoStop}%
\end{thebibliography}%

%\begin{thebibliography}{99}
%\bibitem[label1]{cite_key1} bibliographic information
%\bibitem[label1]{cite_key1} bibliographic information
%\bibitem[label1]{cite_key1} bibliographic information
%\bibitem[label1]{cite_key1} bibliographic information
%\bibitem[label1]{cite_key1} bibliographic information
%\end{thebibliography}

\end{document}